# Synthesis Method-dependent Resist-like Behavior of Liquid Crystalline Graphene Oxide Films Under Electron Beam Irradiation

A. Sahakyan[1], Ghada E. Khedr[2], A. A. Hovhannisyan[3], G. Gevorgyan[4], K. Margaryan[4], G. Shakhatuni[4], M. Aleksanyan[4], Mohamed Aly Saad Aly[5,6], N. Nwaji[7], H. Gharagulyan[1, 4*]

[1] Liquid Crystalline Nanosystems Research Group, A.B. Nalbandyan Institute of Chemical Physics NAS RA, Yerevan 0014, Armenia

[2] Department of Analysis and Evaluation, Egyptian Petroleum Research Institute, Cairo 11727, Egypt

[3] The Scientific Technological Centre of Organic and Pharmaceutical Chemistry NAS RA, Molecule Structure Research Centre, 26 Azatutyan Av., Yerevan, 0014, Armenia

[4] Institute of Physics, Yerevan State University, Yerevan 0025, Armenia

[5] School of Electrical and Computer Engineering, Georgia Institute of Technology, Atlanta, GA 30332, United States

[6] Department of Electrical and Computer Engineering at Georgia Tech Shenzhen Institute (GTSI), Shenzhen, Guangdong 518055, China

[7] Institute of Fundamental Technological Research, Polish Academy of Sciences, Pawinskiego 5B Str. 02-106, Warsaw, Poland

*Corresponding Author, Dr. Hermine Gharagulyan, Email: herminegharagulyan@ysu.am

## Abstract

Understanding irradiation-driven transformations in liquid crystalline graphene oxide (GO) hybrids allows for the precise, chemical-free tuning of material properties. By exposing GO to controlled radiation, researchers can selectively reduce it, tailor its bandgap, and construct multifunctional microstructured devices for energy storage, sensing, and flexible electronics. In particular, electron beam techniques are foundational for understanding and engineering nanomaterials, thus providing atomic-scale resolution for real-time observation of dynamic structural transformations and chemical evolution, while also serving as precise tools for controlled, localized material modification, and nanofabrication. Chitosan-functionalized liquid crystalline GO represents a highly synergistic nanocomposite that merges the mechanical strength and massive surface area of ordered GO with the biocompatibility, film-forming ability, and hydrogen-bonding network of chitosan. Together, they create a non-toxic, versatile, and highly responsive platform for a broad spectrum of advanced applications. Here, the electron-beam-induced structural evolution and patterning behavior of chitosan-functionalized GO films, specifically those using chemically vs. electrochemically exfoliated GO, are investigated. Experimental results (supported by DFT calculations) revealed that electron irradiation drives controllable microstructure formation in GO by selectively triggering ablation and densification pathways, which are dictated by the GO's synthesis route and initial oxidation state. DFT calculations were performed on representative highly oxidized (CGO-like) and moderately oxidized (ECGO-like) bilayer models, including representative chitosan–GO configurations. For CGO, significant C–C bond elongation near oxygen functionalities (1.504 Å vs. 1.420 Å) and a favorable decarboxylation reaction energy (−0.999 eV) support the plausibility of carbon-containing group removal. Incorporation of a chitosan monomer modifies this energetic tendency, with the corresponding decarboxylation energy provisionally estimated to change to −0.42 eV, indicating that chitosan–GO interactions can influence the local carbon-loss energetics. In contrast, ECGO exhibits oxygen-removal-associated interlayer contraction (4.9 → 4.5 Å), accompanied by charge redistribution and enhanced C 2p contributions near the Fermi level, consistent with structural densification and partial restoration of a carbon-rich/sp$^2$-like network. These results provide atomistic insight into the oxidation-state-dependent electron-beam response of liquid-crystalline GO/chitosan systems. Furthermore, a comprehensive characterization study was conducted to compare the optostructural properties of the present system prior and subsequent to electron-beam irradiation. Moreover, a systematic study of the films' mechanical properties across both micro- and macroscopic scales was also performed to elucidate the underlying mechanisms responsible for the observed phenomena. The results demonstrate that the integration of mask-free electron-beam patterning, multiscale characterization, and DFT modeling establishes a previously unexplored synthesis–structure–irradiation relationship in GO films, providing fundamental insights into irradiation-induced microstructure evolution and opening new avenues for the design of GO-based micro- and nanostructured materials.

## 1. Introduction

Graphene oxide (GO) is an extremely versatile, functional nanomaterial; and due to its unique structural flexibility, rich chemical functionality, and extraordinarily high surface area, it can be tailored for a massive array of advanced technological applications [1]. In addition, when exfoliated into high-aspect-ratio nanosheets, GO can spontaneously self-assemble into nematic liquid crystalline (LC) phases at relatively low concentrations, driven by excluded-volume interactions and electrostatic repulsion from negatively charged functional groups [2]. This self-assembly process allows for the creation of macroscopic, highly aligned structures that dramatically improve the mechanical, optical, and transport characteristics of the final films compared to unaligned, isotropic GO films [3].

The exfoliation method used to synthesize GO is the defining step that dictates its final physicochemical properties, such as lateral sheet size, layer thickness, defect density, and the existence of oxygen-containing functional groups. Chemically exfoliated graphene oxide (CGO) is a very oxidized, rich in structural defects, tends to form crumpled, relatively smaller sheets, and exhibits pronounced hydrophilicity due to abundant oxygen-containing surface functional groups [4]. In contrast, electrochemically exfoliated graphene oxide (ECGO) is a greener, more scalable alternative to harsh chemical methods. By using electrical potential to intercalate ions and expand graphite layers, it produces larger, flatter flakes with fewer basal-plane defects and a higher preservation of conductive networks and sp$^2$ domains [5]. These differences govern its LC phase behavior, electrostatic interaction with chitosan, film microstructure, and mechanical performance.

Chitosan (Chit) is an ideal biopolymer for functionalizing GO owing to its synergistic integration of reactive functional groups (amine and hydroxyl), non-toxic biocompatibility, excellent film-forming properties, and intrinsic chirality [6]. Chit-GO nanocomposites offer a powerful synergistic platform. The amino and hydroxyl groups of Chit readily form hydrogen bonds and electrostatic interactions [7] with GO's oxygen-containing groups, improving colloidal stability and mechanical stiffness. This structural cross-linking restricts polymer swelling, limits aggregation, and induces chirality transfer [8], thus enabling highly selective molecular recognition. Such composites show significant promise across multiple applications ranging from sensing [9] and drug delivery [10] to antifouling of ultrafiltration membranes [11] and protection against enzymatic cleavage [12].

Chitosan-functionalized graphene oxide (Chit-GO) films utilize the excellent film-forming and biocompatible traits of Chit combined with the structural strength of GO. They are primarily synthesized using top-down exfoliation (chemical or electrochemical) of graphite, followed by solution-based fabrication (like drop casting, spin coating, or wet spinning from both LC and non-LC GO dispersions), followed by drying or cross-linking to optimize mechanical integrity [7, 13, 14]. In addition, Chit-functionalization tailors interactions between GO sheets by balancing repulsive forces and attractive hydrogen bonding. This promotes stable lyotropic LC phases at specific concentrations, which act as highly ordered templates for film casting and wet-spinning processes [15].

Scanning electron microscopy (SEM) goes beyond passive observation. By leveraging focused, high-energy electron beams, physical and chemical changes can be actively induced to materials inside the vacuum chamber in real time, transforming the SEM into a dynamic, nanoscale fabrication and testing tool [16, 17]. Literature revealed that electron beam (e-beam) irradiation relies on the formation of very reactive free radicals, which subsequently drive radiolytic degradation processes such as deoxygenation [18], deamination [19], and depolymerization [20], inducing phase transitions of GO layers into amorphous carbon depending on beam energy and dose [21]. Therefore, such technique can achieve scalable production of reduced graphene oxide (rGO) without using hazardous chemicals [22]. Furthermore, it allows for precise surface modification and micro/nanopatterning without the need for complex mask-based procedures [23]. Christman *et al.* detailed a versatile electron beam lithography technique used to precisely organize and position multiple proteins with nanoscale precision [24].

This study investigated how different GO exfoliation routes impact the morphology, LC behavior, mechanical strength, and e-beam patterning capabilities of Chit-functionalized GO composite films. In particular, chemically and electrochemically exfoliated GO-based composites were compared to establish guidelines for advanced material design with tunable structural and mechanical properties. The comparison of these composites across key structural and mechanical parameters (such as LC domain formation, macroscopic alignment, chirality, tensile strength, and hardness) highlights specific differences that dictate their use in multifunctional applications. In addition, e-beam-induced ablation and densification mechanisms in LC GO films prepared via chemical and electrochemical exfoliation were experimentally and theoretically investigated, highlighting the strong dependence of resist-like behavior on the synthesis method. The novelty of this work lies in demonstrating that the electron-beam-induced structural and morphological responses of GO-based films can be controlled not only by irradiation parameters but also by the chemical nature of the starting GO and the incorporation of chitosan, resulting in distinct irradiation-induced patterns, including holes and protrusions.

## 2. Reagents and Methodology

### 2. 1. Reagents

Chitosan (>75%, 300-375 kDa), Graphite foil, Acetic Acid (Hac, $CH_3COOH$), Potassium Permanganate ($KMnO_4$), Ammonium sulfate ($(NH_4)_2SO_4$), Sulfuric acid (98%, $H_2SO_4$), Phosphoric acid (85%, $H_3PO_4$), and Hydrogen peroxide (30%, $H_2O_2$) were acquired from Sigma-Aldrich. All solvents and chemicals were purchased at a high-purity commercial standard and were used exactly as received without performing additional purification steps in the laboratory. In all experiments, ultrapure deionized water (18 MΩ·cm) was used at 25°C.

### 2. 2. Chitosan-Functionalization of LC GO and Film Preparation

The synthesis and the LC phase formation of GO followed previously reported works [21, 25]. The preparation of Chit-GO composites was performed as follows: 15 mg of GO powder obtained by electrochemical and chemical synthesis was dissolved in 15 mL of water and treated with mild ultrasonication for 30 minutes. Next, 150 mg of Chit powder and 0.3 mL of HAc were sequentially added and poured into each solution under stirring. After mild ultrasonication for 2 h, the Chit-GO solutions were heated up and simultaneously stirred at 90 °C for 10 hours. Upon completion, 100 µL, 200 µL, and 1 mL of each suspension were cast onto glass plates and dried out under ambient conditions at 25 °C for approximately 36 hours, yielding films with diameters of approximately 0.5, 1, and 5 cm, correspondingly. The resulting dry films were then carefully peeled off the glass plates. **Scheme 1** shows a schematic illustration of the functionalization procedure of LC GO with chitosan, and the film preparation.

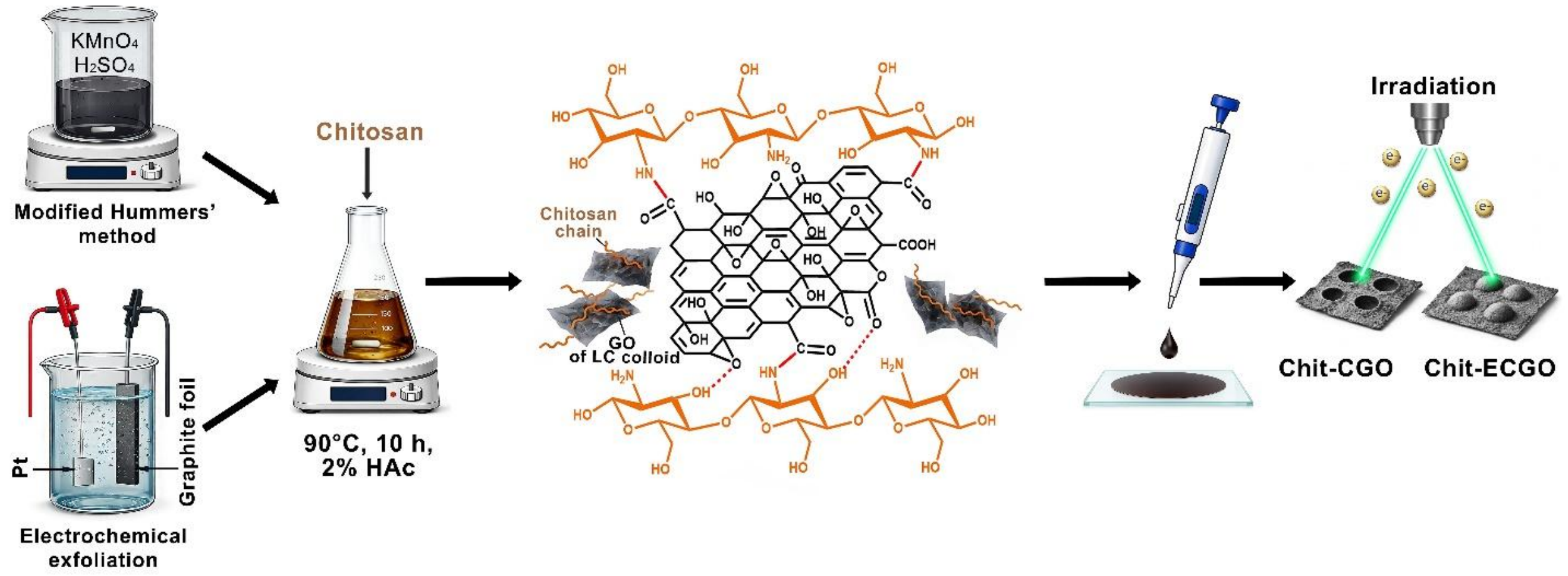

**Scheme 1**. Schematic diagram of the Chit- LC GO film fabrication and e-beam irradiation process.

### 2. 3. Instrumentations

The optical properties of the samples were analyzed utilizing an Agilent Cary 60 ultraviolet-visible (UV–Vis) spectrophotometer (Agilent Technologies, Santa Clara, CA, USA). The structural organization and chiral ordering of the Chit-CGO and Chit-ECGO samples were investigated by circular dichroism (CD) spectroscopy (DSM 20, Olis, USA). The Thermo Scientific K-Alpha XPS System is a highly surface-sensitive X-ray Photoelectron Spectroscopy and was utilized to quantify the facial elemental composition and identify the local chemical environments, oxidation states, and valence states of atoms. Chemical bonding environments and carbon hybridization states were recognized through Raman spectroscopic measurements taken by the Bruker SENTERRA II Confocal Microscope (Brucker, Germany). FTIR-ATR spectroscopy using the PerkinElmer Spectrum II, a highly effective, non-destructive method, was used for characterizing chemical bonding. The Thermo Scientific Prisma E SEM, an advanced, flexible Scanning Electron Microscope (SEM), was utilized to analyze the surface morphology, topography, and shape of the synthesized materials. Instec HCS350G-TNS tensile testing system was utilized to assess the mechanical activities of the fabricated samples. Hardness measurements of the Chit-GO films were performed using a microhardness tester (FALCON 600G2FA, INNOVATEST Europe BV, The Netherlands).

### 2. 4. Computational details

Density functional theory (DFT) calculations were performed using the Vienna Ab initio Simulation Package (VASP 5.4) to elucidate plausible atomistic factors underlying the distinct e-beam-induced responses of chemically and electrochemically exfoliated GO films. The exchange-correlation interactions were described within the generalized gradient approximation (GGA) using the Perdew-Burke-Ernzerhof (PBE) functional. The projector augmented-wave (PAW) method was employed to describe the interaction between valence electrons and ionic cores. A plane-wave kinetic energy cutoff of 450 eV and convergence criteria of $10^{-5}$ eV for electronic self-consistency and 0.02 eV $Å^{-1}$ for atomic forces were used throughout the calculations.

Two bilayer GO models were constructed as simplified representations of the highly oxidized CGO ($C_{64}H_7O_{14}$) and moderately oxidized ECGO ($C_{64}H_5O_{10}$) systems [26,27]. These compositions were chosen to represent relative oxidation levels, not to reproduce the exact structure of the experimental films. To explore the two experimentally observed responses, we considered (i) a hole-defect configuration for CGO, in which a carbon vacancy and neighboring oxygen groups were removed, and (ii) a reduced, contracted (hump-like) configuration for ECGO, in which selected oxygen groups were removed without introducing carbon vacancies. We emphasize that these configurations were chosen based on the experimentally observed morphologies; the calculations therefore evaluate the energetic and electronic consequences of these assumed pathways and do not independently predict which pathway is favored. To assess competing pathways, the carbon-loss (decarboxylation) and oxygen-removal reactions were evaluated for both models. All structures were fully optimized until the residual atomic forces were below the convergence threshold. The optimized geometries, total energies, charge distributions, and electronic structures were subsequently analyzed to evaluate the relative stability of the two irradiation-induced transformation pathways.

To examine the role of chitosan, a single chitosan monomer (glucosamine unit) was placed on the upper GO layer of both models in initial configurations, and the lowest-energy structure was selected. The adsorption energy was calculated as $E_{ads} = E(GO+chitosan) - E(GO) - E(chitosan)$. The same optimization and analysis procedures were then applied. Atomic charges were analyzed using Bader charge analysis, and the PDOS was calculated for the ECGO and hump structures.

## 3. Findings and Discussion

### 3. 1. Structural characterization of Chitosan-functionalized GO Films

**Figure 1.** shows the characteristic spectral profile of CGO and ECGO. The hallmark optical absorption signatures of CGO are presented in the UV-Vis spectrum exhibited in **Figure 1(a).** Electronic transition peaks at about 232 nm are accredited to π-π* conversions from the aromatic C=C bonds, and at about 300 nm, corresponded to n-π* conversions of C=O bonds [28]. A similar spectral profile is observed for ECGO, with characteristic bands located at around 237 nm, and 295 nm [21]. The UV-Vis spectra of the Chit-CGO and Chit-ECGO composites, presented in **Figure 1(b)**, exhibit absorption maxima at 251 nm and 258 nm, respectively. The shift in these absorption maxima could be ascribed to the interactions between Chit and GO [8].

The Circular Dichroism (CD) spectra of CGO, ECGO, and their corresponding composites are presented in **Figure 1(c, d)**. Chit is CD-active molecule due to its chiral polysaccharide backbone and typically exhibits a broad negative band in the 190-230 nm region, arising from overlapping n-π* and π-π* electronic modifications of the carbonyl groups (C=O) present in Chit [29]. In contrast, CGO does not show a detectable CD signal under the studied conditions, whereas ECGO exhibits a negative CD band at approximately 240 nm. It is pertinent to state that pristine planar graphene is achiral. The observed CD signal in ECGO indicates that the electric field driven process led to localized intercalation of ions and mechanical stress which result in asymmetric cracking and formation of in-plane ripples, hence chiral conformation in the carbon base plane [30]. After functionalization with Chit, the CD spectra of the composites display negative bands at 228 nm for Chit-CGO and 230 nm for Chit-ECGO. The positions of these bands differ from those reported for the individual components, indicating changes in the electronic transitions within the system. These changes are attributed to interactions between Chit and GO, which modify the local chemical environment of the relevant functional groups.

For XPS studies of Chit-CGO and Chit-ECGO nanocomposites, initially survey spectra carried out to acquire a wide-ranging spectrum scan as shown in **Figures 1(e, i**). Here, it was found that the O/C ratio to be 0.55 and 0.43 for Chit-CGO and Chit-ECGO nanocomposites, individually, which was calculated according to previously reported study [31]. **Figures 1(f-h)** and **2(j-l)** exhibit the fitted and initial spectra for the elemental composition of the Chit-CGO and Chit-ECGO samples, specifically, O1s, N1s and C1s, correspondingly. The XPS C1s core-level spectrum of the Chit-CGO composite shows three peaks at positions 287.6, 285.2, and 284.0 eV, which correspond to –C=O, C=O, and C–C/C=C, individually. In analogy, the C1s spectrum of the Chit-ECGO composite likewise exhibits 3 different peaks sited at binding energies 287.9, 286.3, and 284.5 eV, which are attributed to –C=O, C=O, and C–C/C=C separately. The binding energy signatures for Chit-CGO and Chit-ECGO composites are represented in **Table 1**.
The positive shift in binding energy of C 1s and O 1s in Chit-ECGO relative to Chit-CGO could be attributed to existence of carbon atoms in less defective networks in Chit-ECGO due less harsh preparation condition [32]. On the other hand, the N 1s of Chit-ECGO show a negative shift in binding energy compared to that of Chit-CGO. This can be explained on the fact that the use of electrolyte significantly lower harsh oxidizing agent and inherently lead to the preservation of graphitic sp2 carbon lattice which enhances electron density around N atom and effectively lowers the core-level binding energy [33].

**Table 1.** XPS binding energy signatures for Chit-CGO and Chit-ECGO composites.

| Chit-CGO | | | Chit-ECGO | | |
|---|---|---|---|---|---|
| *Energy signature (eV)* | | | *Energy signature(eV)* | | |
| C1s | $sp^3C$<br>C–OC, C–OH<br>O–C=O | 284.0<br>285.2<br>287.6 | C1s | $sp^3C$<br>C=O<br>O–C=O | 284.5<br>286.3<br>287.9 |
| O1s | | 531.8 | O1s | | 532.2 |
| N1s | | 399.0<br>401.4 | N1s | | 398.9<br>400.7 |

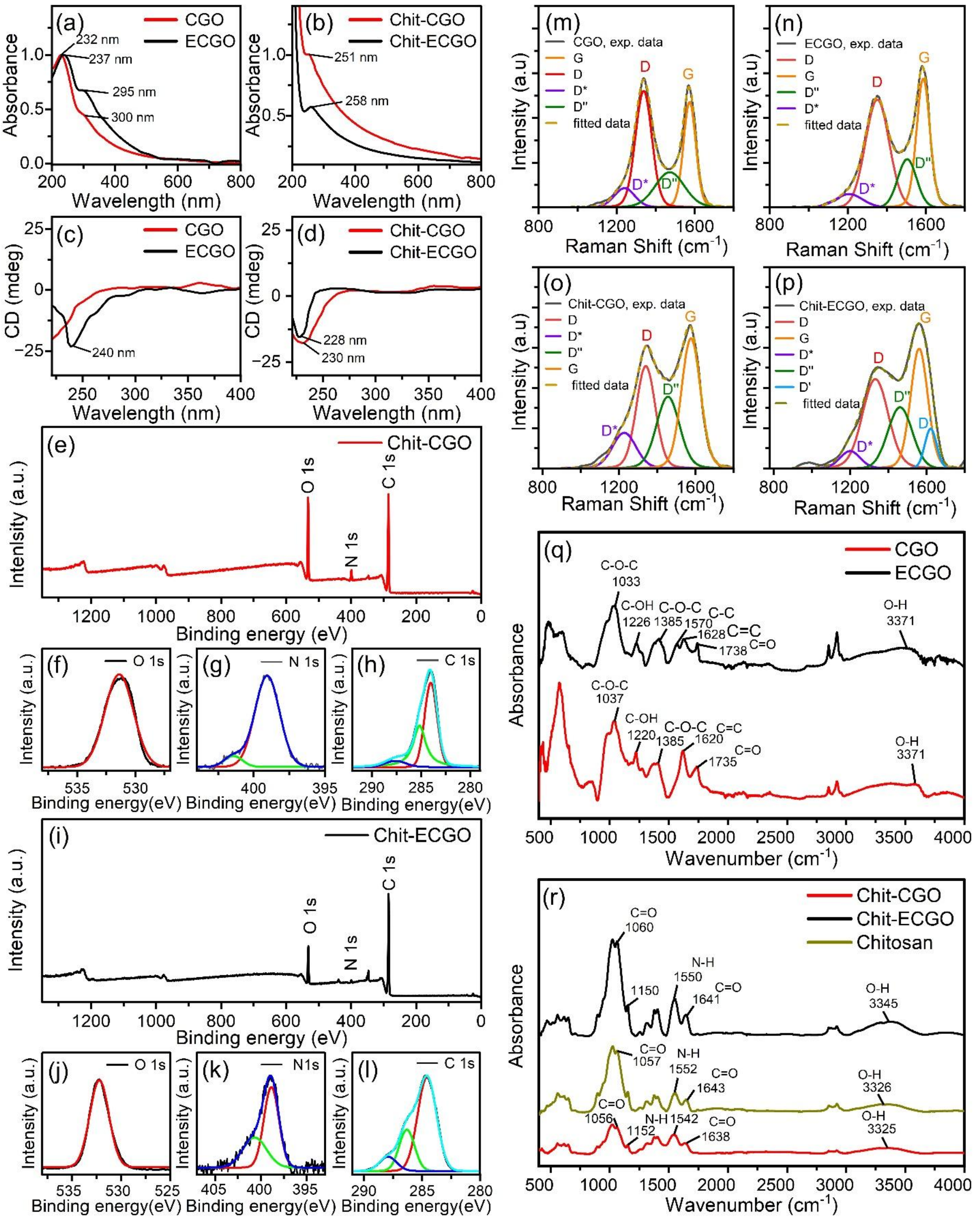
(a)
CGO
ECGO
232 nm
237 nm
295 nm
300 nm
Absorbance
Wavelength (nm)
(b)
Chit-CGO
Chit-ECGO
251 nm
258 nm
Absorbance
Wavelength (nm)
(c)
CGO
ECGO
240 nm
CD (mdeg)
Wavelength (nm)
(d)
Chit-CGO
Chit-ECGO
228 nm
230 nm
CD (mdeg)
Wavelength (nm)
(e)
Chit-CGO
O 1s
N 1s
C 1s
Intenisity (a.u.)
Binding energy (eV)
(f)
O 1s
(g)
N 1s
(h)
C 1s
Intensity (a.u.)
Binding energy(eV)
(i)
Chit-ECGO
O 1s
N 1s
C 1s
Intensity (a.u.)
Binding energy (eV)
(j)
O 1s
(k)
N1s
(l)
C 1s
Intensity (a.u.)
Binding energy(eV)
(m)
CGO, exp. data
G
D
D*
D''
fitted data
Intensity (a.u)
Raman Shift (cm-1)
(n)
ECGO, exp. data
D
G
D''
D*
fitted data
Intensity (a.u)
Raman Shift (cm-1)
(o)
Chit-CGO, exp. data
D
D*
D''
G
fitted data
Intensity (a.u)
Raman Shift (cm-1)
(p)
Chit-ECGO, exp. data
D
G
D*
D''
D'
fitted data
Intensity (a.u)
Raman Shift (cm-1)
(q)
CGO
ECGO
C-O-C 1033
C-OH 1226
C-O-C 1385
C-C 1570
1628
C=C
1738
C=O
O-H 3371
C-O-C 1037
C-OH 1220
C-O-C 1385
C=C 1620
1735
C=O
O-H 3371
Absorbance
Wavenumber (cm-1)
(r)
Chit-CGO
Chit-ECGO
Chitosan
C=O 1060
1150
N-H 1550
C=O 1641
O-H 3345
C=O 1057
N-H 1552
C=O 1643
O-H 3326
C=O 1056
1152
N-H 1542
C=O 1638
O-H 3325
Absorbance
Wavenumber (cm-1)

**Figure 1**. Characterization of CGO, Chit-CGO, ECGO and Chit-ECGO nanocomposites. (a) UV-Vis absorption and (c) CD spectra for the CGO and ECGO. (b) UV-Vis absorption and (d) CD spectra for the Chit-CGO and Chit-ECGO (red and black traces, respectively); (e, i) XPS survey scan spectra: (f), (j) O1s; (g), (k) N1s and (h), (l) C1s peaks for Chit-CGO and Chit-ECGO nanocomposites. The C1s main peak of carbon tape at 284 eV was used as the XPS energy reference for all binding energies; (m-p) Raman spectra deconvolution and peaks fitting for: (m) CGO; (n) ECGO; (o) Chit-CGO; (p) Chit-ECGO. A 532 nm laser line was focused via a 50× magnifying lens. A diffraction grating with 1200 grooves per mm was utilized. A 2000–500 $cm^{-1}$ of spectral window was designated, and a 6.25 mW laser was kept for 10 seconds, delivering for all of the probes. (q, r) FT-IR spectra: (q) for CGO and ECGO (red and black traces, respectively). (r) Chit-CGO Chit-ECGO and Chitosan (red, black and green traces, respectively). All measurements were carried out at room temperature.

In addition, **Figures S1(a-j)** present the XPS results of pristine Chit, CGO, and ECGO. Dual overlapping peaks in the N 1s spectra exhibited in **Figure S1(i)** are for Chit. The peak at 400.5 eV is attributed to amide and amine (primary and tertiary) groups, while the peak at 401.8 eV, is correlated to protonated amine ($–NH_3+$) species [34].
If the carboxyl groups of GO react with the amino groups of Chit, the area ratios (*R*) of these peaks will rise progressively due to the increase in amount of amide groups. The corresponding *R*s are summarized in **Table S1**. For the Chit-CGO composite, the N 1s spectrum shows dual peaks at 399.0 and 401.4 eV, appointed to amine/amide nitrogen and protonated amine species, individually. The corresponding intensity ratio increases significantly to 3.25. Similarly, the N 1s spectrum of the Chit-ECGO composite shows two components located at 398.9 and 400.7 eV, credited to amine/amide and protonated amine groups, correspondingly, with an *R* value of 1.70 [35].

A quantitative XPS comparison of the O/C ratio and deconvoluted functional groups for CGO, ECGO, Chit-CGO, and Chit-ECGO is provided in **Table S2** in the Supplementary Materials.

Detailed analysis of Raman spectra provides both qualitative and quantitative information regarding crystallite size, structural disorder, the number of layers, and defects [36]. The Raman spectra of GO fabricated *via* chemical and electrochemical routes are presented in **Figures 1(m)** and **1(n)**, respectively. They were formed from 10 diverse spatial positions of each sample. In graphene-derived materials, the first-order Raman spectrum is typically dominated by the *D* and *G* bands. Numerous studies have demonstrated that the Raman intensity ratio, $I_D/I_G$, is inversely correlated to the in-plane $sp^2$ crystallite domain size, indicating that an increase in defect density results in smaller graphitic domains. However, it is worth noting that Raman response does not hold universally for highly disordered carbon materials. To obtain accurate intensity values, the first-order Raman spectra of the two samples were simplified and fitted using 4 primary components and for CGO: the *D* band (~1343 $cm^{-1}$), the *D″* band (~1472 $cm^{-1}$), the *G* band (~1581 $cm^{-1}$) and the *D** band (~1240 $cm^{-1}$). The *D* band, placed at nearly 1343 $cm^{-1}$, originates from the breathing modes of $sp^2$ carbon rings and is trigered by the presence of structural defects. The *G* band, observed near 1581 $cm^{-1}$, corresponds to the in-plane stretching vibration (*E2g* mode) of $sp^2$-bonded carbon atoms. The intensity ratio of the *D* and *G* bands ($I_D/I_G$) in the first-order Raman spectrum is the primary spectroscopic metric used to define the in-plane crystallite size (*La*) of $sp^2$ domains in GO nanoparticles [37, 38]. The $I_D/I_G$ ratio was determined from the fitted peak intensities and the findings are presented in **Table S3**:

The same four primary components observed for ECGO are: the *D* band (~1350 cm⁻¹), the *D″* band (~1495cm⁻¹), the *G* band (~1585 cm⁻¹), and the *D** band (~1200 cm⁻¹). Notably, the $I_D/I_G$ ratio of CGO (1.56) is higher than that of ECGO (1.32), demonstrating a superior level of structural disorder in the chemically synthesized sample as shown in **Figure S2**. Since the $I_D/I_G$ ratio is inversely correlated with the $sp^2$, the higher value observed for GO suggests smaller $sp^2$ domain sizes and a higher defect density. This increase in disorder may also be associated with enhanced oxidation levels and a greater concentration of oxygen-containing functional groups [40].

The CGO sample exhibits a higher $I_D/I_G$ ratio relative to ECGO, indicating a greater degree of structural disorder and increased defect density and a higher degree of disruption in the $sp^2$ carbon network [3]. Conversely, the slightly lower $I_D/I_G$ ratio observed for ECGO implies a comparatively higher level of crystallinity and larger $sp^2$ domain sizes than those of CGO. These findings prove that the synthesis route significantly influences structural ordering and defect distribution in the resulting graphene-based materials.

Interestingly, after Chit-functionalization, the Raman bands of both GO samples behaved similarly: the $I_D/I_G$ ratio decreased (**Figures 1(o) and 1(p)**). Very similar results were obtained for polydopamine-functionalized GO, where after functionalization, the Raman spectra presented two identical GO bands (*D* and *G*) in the two samples, with the $I_D/I_G$ ratio for Chit-GO being lower than that of GO, indicating different crystalline properties and defects, as well as lower oxygen content [41]. Following functionalization, noticeable shifts in the locations of the *D″* and *D** bands were noted: the *D″* band shifted toward lower wavenumbers, whereas the *D** band shifted toward higher wavenumbers. The positions of these bands may serve as sensitive indicators for evaluating the degree of GO functionalization. More recently, density functional perturbation theory (DFPT) simulations have provided further insight into the basis of the *D** band in graphene oxide systems [42]. These calculations show that the *D** band arises from vibrational modes of carbon atoms whose motion is constrained by the existence of oxygen-containing functional groups. It was also shown that the peak locations of the *D″* and *D** bands display a distinct dependence of the content of oxygen in GO and, thus, can be utilized as good factors to approximate the content of oxygen [38]. The results of the Raman depth-profiling experiments for the Chit-CGO and Chit-ECGO are presented in **Figure S3**.

The FT-IR spectrum of CGO displays prominent peaks at 1735, 1620, 1385, 1220, and 1037 $cm^{-1}$ for C=O stretch in carboxylic, C=C aromatic stretch, C−O−C ether, C−OH stretch in acids, and C−O−C epoxy groups, correspondingly [5]. In the case of ECGO the absorption bands observed at 1737, 1628, 1385, 1226, and 1033 $cm^{-1}$ are credited to the stretching vibrations of C=O carbonyl, C=C aromatic, C−O−C ether, C–OH carboxylic, and C–O–C epoxy functional groups, individually (**Figure 1(q)**). The peak at 1570 $cm^{-1}$ is correlated with the skeletal stretching of the C–C alkene group. The pronounced intensity of the characteristic bands in both spectra indicates a high abundance of oxygen-containing functional groups following the oxidation process. The Chit-CGO spectrum demonstrates a band at 1638 $cm^{-1}$ owing to the stretching vibration of C=O in the amide developing from GO carboxyl groups (COOH), Chit amine ($NH_2$) groups overlapping with the amine I band of Chit (**Figure 1(r)**). Also, there is an absorption band at the 1152 $cm^{-1}$, assigned to the principal alcoholic group of $C_6$–OH. The band at 1542 $cm^{-1}$ relates to –N–H bending of secondary amides subsequent to creating the bond with CGO [43]. Analogous bands are likewise observed in the Chit-ECGO spectrum, exhibiting slight shifts in their respective wavenumbers. Overall, these spectral features in both systems support covalent functionalization between the carboxylic groups of GO (both chemically and electrochemically derived) and the amino groups of Chit with slight differences in peak positions arising from structural and compositional variations between CGO and ECGO.

## 3. 2. Mechanical Properties of Chitosan-functionalized CGO and ECGO Films

Tensile strength and hardness measurements of pristine and Chit-functionalized CGO and ECGO films were systematically investigated. The mechanical property measurements were performed using 10 independent samples, and the results are presented as the mean ± standard deviation. To determine the hardness parameters of the prepared pristine Chit, CGO, ECGO, Chit-CGO and Chit-ECGO prepared films, the Vickers hardness measurement method was employed. Specifically, five random points on the surface of each film were selected for analysis. The measurements were performed using the HV 0.5 hardness scale with a dwell time of 5 seconds. **Figure 2** graphically presents the measurement data together with the calculated average hardness values for the studied films. The detailed measurement results are presented in **Table S4**. The ECGO film exhibits a hardness value approximately one-fold larger than that of the corresponding CGO film. CGO commonly contains vacancy defects, oxidative holes, and amorphous carbon regions, which reduce mechanical strength and hardness. The hardness depends on the balance between structural integrity of $sp^2$ domains, defect type, sheet size, interlayer packing, and film morphology. On the contrary, the hardness values of Chit-functionalized CGO and ECGO differ only slightly. This reduced difference suggests that the mechanical response becomes increasingly governed by polymer-GO interfacial interactions and composite network formation, rather than solely by the intrinsic structural differences of the GO sheets [44].

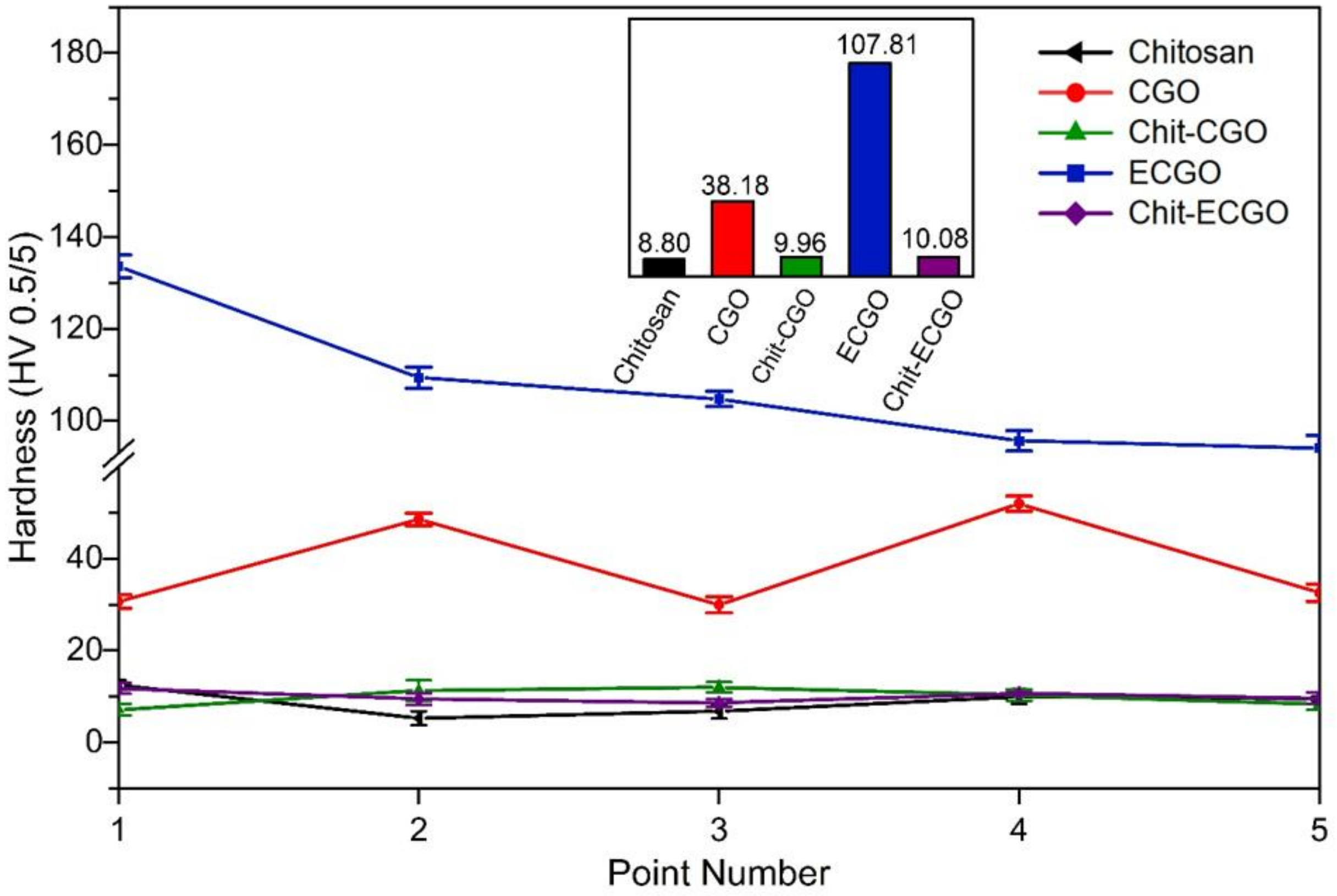

**Figure 2.** Hardness diagram measured at five selected surface locations of Chit, CGO, ECGO, Chit-CGO, and Chit-ECGO films, together with the corresponding average hardness values.

In parallel, tensile strength measurements were also performed for the above-mentioned films. The stress-strain curves of pristine Chit, CGO, ECGO, and Chit-functionalized CGO and ECGO films at different strain rates are depicted in **Figures 3(a)** and **3(b)**.

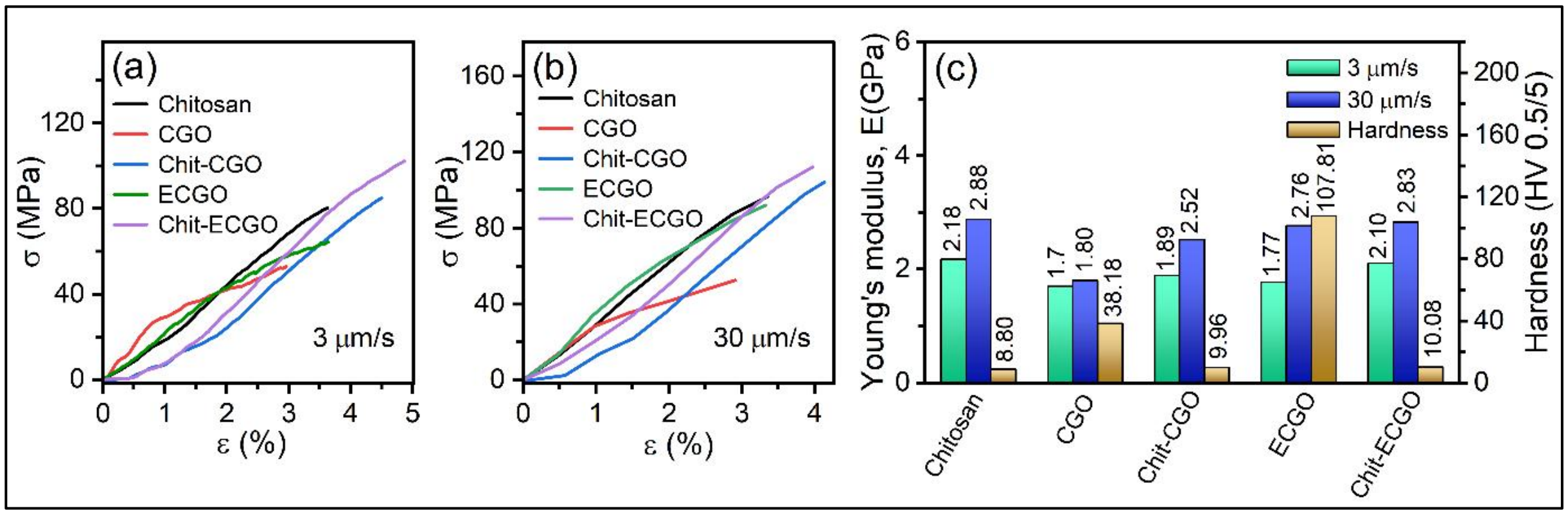

**Figure 3.** Stress-strain curves of Chit, CGO, ECGO, Chit-CGO, and Chit-ECGO films measured at strain rates of (a) 3 μm/s and (b) 30 μm/s. (c) Comparative diagram of Young's modulus and hardness for the GO-based films. The standard deviations for the samples at strain rate of 3 μm/s are 0.92, 0.61, 0.90, 0.74, and 1.27 for Chitosan, CGO, Chit-CGO, ECGO, and Chit-ECGO,

respectively, while those at strain rate 30 μm/s for the samples are 1.20, 0.62, 1.13, 1.20, and 1.50, respectively.

Tensile testing revealed that pristine GO films exhibited relatively poor mechanical performance. CGO displayed Young's moduli of 1.70 and 1.80 GPa at strain rates of 3 and 30 μm/s, respectively, whereas ECGO exhibited values of 1.77 and 2.76 GPa under the same conditions. Chit functionalization enhanced the mechanical performance of both materials. The Young's modulus increased to 1.89 and 2.52 GPa for Chit-CGO and to 2.10 and 2.83 GPa for Chit-ECGO at strain rates of 3 and 30 μm/s, respectively. This improvement is attributed to strengthened interfacial interactions between GO nanosheets and Chit, which promote more efficient load transfer and enhanced structural cohesion within the composite films. **Figure 3(c)** presents a comparative analysis of Young's modulus and hardness for Chit, CGO, ECGO, Chit-CGO, and Chit-ECGO films. As observed, the incorporation of chitosan increases the stiffness of the Chit-GO composites, accompanied by a reduction in their elastic deformability [45].

### 3.3 Electron-Beam Micropatterning of Chitosan-modified GO Films

The changes in the facial morphology of Chit-modified CGO and ECGO were investigated by SEM under the subsequent settings: accelerating potential: 20, 10, 5, and 2 kV; beam current: 4.5nA, 1.2 nA, 0.6 nA, and 100 pA; spot size: 6.0, 5.0, 4.5, and 3.5; scan frame: 768 × 512; pixel size: 0.4 Mpx; dwell time: 4 μs. **Figures 4(a, b)**, **Figures 4(d, e)** and **Figure 4(g, h)** show the morphology and cross-section of Chit-CGO, Chit-ECGO nanocomposites and Chitosan earlier to the e-beam irradiation, respectively. **Figures 4(c), 4(f)** and **4(i)** present the observed outcome of the dynamic processes in Chit-CGO, Chit-ECGO nanocomposites and Chitosan, respectively, subsequent to the e-beam irradiation.

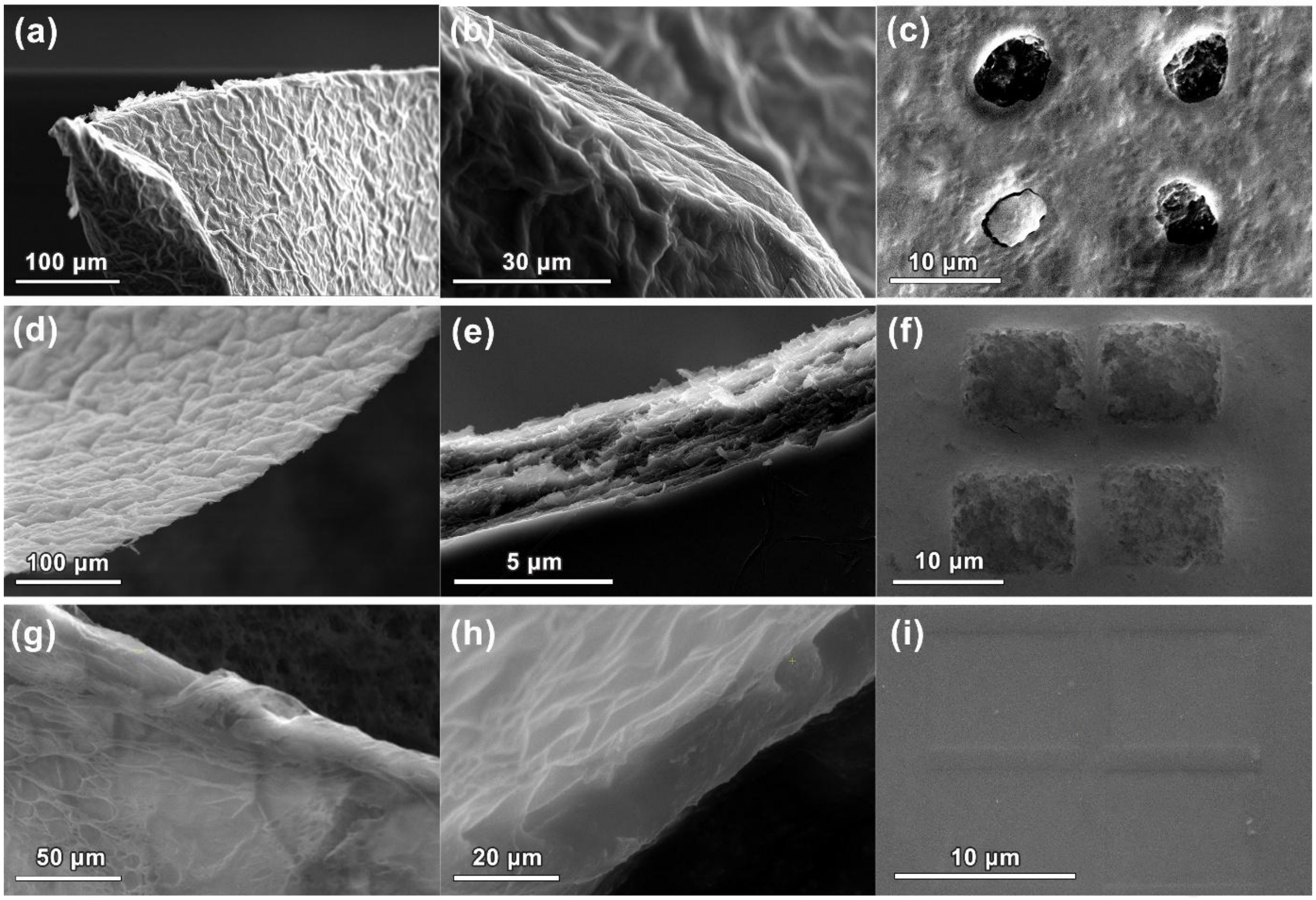

**Figure 4.** SEM images of Chit-CGO, Chit-ECGO and Chitosan morphology **(a, d, g)**, cross section **(b, e, h)** and micropatterns **(c, f, i)**, respectively.

E-beam exposure was quantified by dose. A dose/fluence comparison table for Chit-CGO, Chit-ECGO, and chitosan at an accelerating voltage of 20 kV and a beam current of 0.31 nA is included in **Table 2**, while a separate table for the 10 and 5 kV accelerating voltages are provided in the Supplementary Materials (**Table S5**).

**Table 2.** Absorbed irradiation doses and fluences by Chit-CGO, Chit-ECGO and Chitosan for 20 kV and 0.31 nA.

| System | Irradiation | Area ($cm^2$) | Time (min) | | | | |
|---|---|---|---|---|---|---|---|
| | | | 1 | 2 | 3 | 5 | 10 |
| Chit-CGO | ***Dose ($A{\cdot}s/cm^2$)*** | $5.01\times10^{-8}$ | 0.37 | 0.74 | 1.11 | 1.85 | 3.71 |
| | ***Fluence (electron/$cm^2$)*** | | $2.32\times10^{18}$ | $4.64\times10^{18}$ | $6.96\times10^{18}$ | $1.16\times10^{19}$ | $2.32\times10^{19}$ |
| Chit-ECGO | ***Dose*** | $5.93\times10^{-8}$ | 0.31 | 0.63 | 0.94 | 1.57 | 3.14 |
| | ***Fluence*** | | $1.96\times10^{18}$ | $3.92\times10^{18}$ | $5.88\times10^{18}$ | $9.80\times10^{18}$ | $1.96\times10^{19}$ |
| **Chitosan** | ***Dose*** | $4.69\times10^{-8}$ | 0.40 | 0.80 | 1.19 | 1.98 | 3.96 |
| | ***Fluence*** | | $2.48\times10^{18}$ | $4.96\times10^{18}$ | $7.44\times10^{18}$ | $1.24\times10^{19}$ | $2.48\times10^{19}$ |

Further, the hole size, protrusion height and feature dimensions were analyzed as functions of the e-beam dose and film thickness (**Table S6**).

E-beam irradiation induces markedly different structural responses in Chit-functionalized chemically and electrochemically exfoliated GO films, reflecting differences in their microscopic composition and defect structure. In CGO-based films, the elevated density of oxygen functional groups, namely epoxy, hydroxyl, and carboxyl promote radiolytic bond scission and decarboxylation under electron exposure. The resulting gas evolution ($CO/CO_2$), coupled with Chit chain fragmentation and limited charge dissipation due to low conductivity, leads to localized mass loss and material ejection [46]. This ablation-dominated regime produces well-defined depressions, analogous to "positive-tone resist" behavior, as shown in **Figure 4(c)**. In contrast, Chit-functionalized ECGO possesses a lower oxidation degree, larger $sp^2$ domains and higher electrical conductivity, which suppress extensive radiolytic decomposition and facilitate efficient charge redistribution. Under identical irradiation conditions, electron exposure instead promotes partial reduction of GO, enhanced $\pi$–$\pi$ stacking, and interfacial crosslinking with Chit [18]. Within LC architecture, this reduction-driven reorganization results in interlayer contraction and anisotropic stress redistribution rather than material removal [3, 47]. The confined lateral shrinkage of aligned sheets generates localized out-of-plane protrusions shown in **Figure 4(f)**), giving rise to a compaction-dominated regime reminiscent of "negative-tone resist" behavior [21]. Thus, the e-beam response is determined by the oxidation-dependent balance between fragmentation-driven ablation and reduction-induced densification, revealing a synthesis-controlled transition between two distinct patterning mechanisms in GO-based soft nanostructures, as further supported by the DFT calculations presented below.

XPS and Raman analysis after the e-beam irradiation are illustrated in **Figure 5**. Peak positions for Chit-CGO and Chit-ECGO composites changes slightly after e-beam irradiation. The spectra in **Figures S3 (a)** and **S3(c)** correspond to the Raman depth profiles of Chit-CGO and Chit-CGO irradiated, respectively. In both spectra, the intensity of the *D* and *G* bands decrease proportionally as the focal plane moves deeper into the layer, up to depth of 20 µm. Comparable results were obtained for the Chit-ECGO irradiated samples, as shown in **Figures S3 (c)** and **S3(g)**. The binding energy signatures for Chit-CGO and Chit-ECGO composites are represented in **Table 3**.

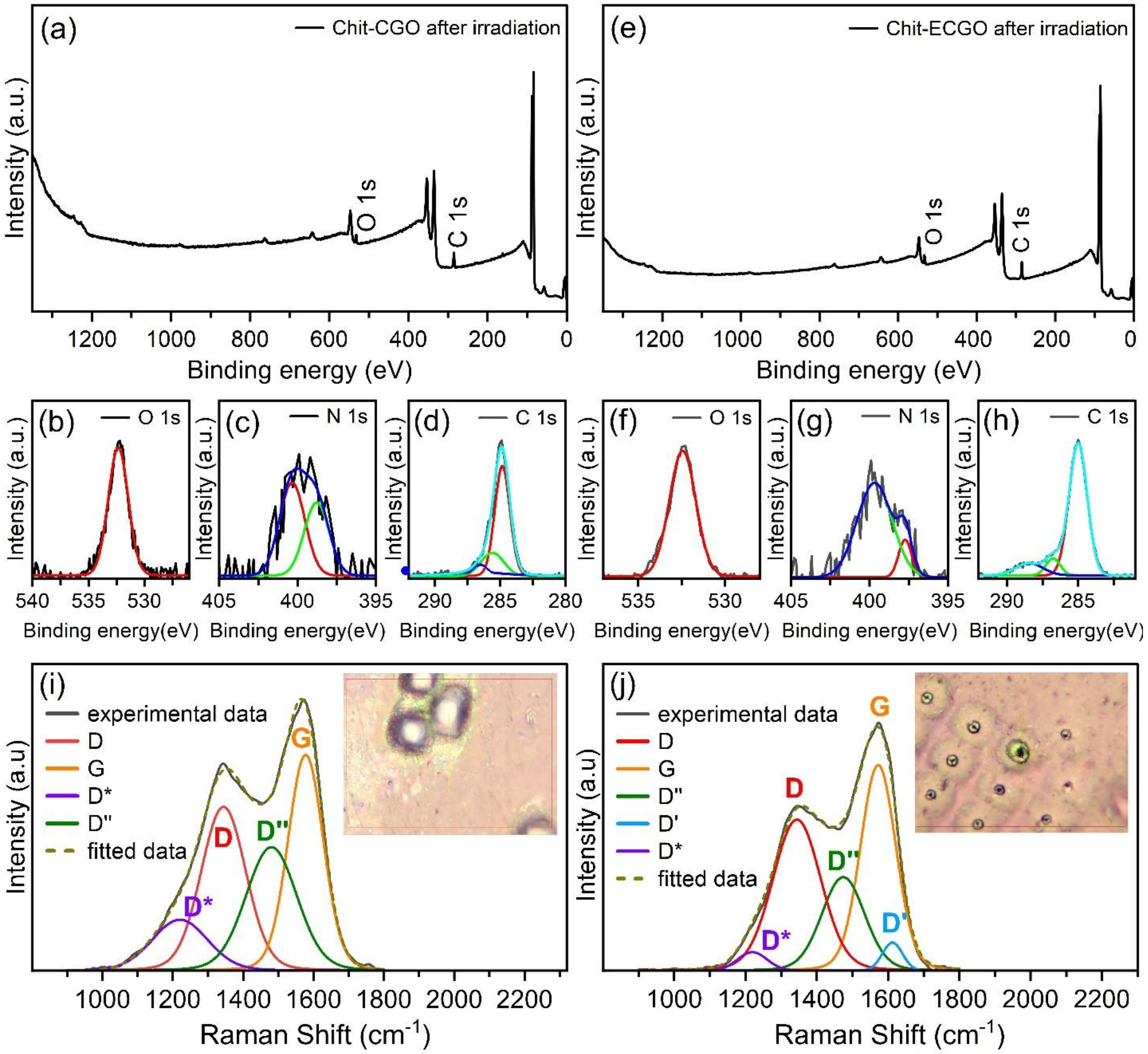

**Figure 5.** Characterization of Chit-CGO and Chit-ECGO nanocomposites after e-beam irradiation: XPS (a-d) and (e-h); Raman (i) and (j) spectra, respectively.

**Table 3.** XPS binding energy signatures for Chit-CGO and Chit-ECGO composites after e-beam irradiation.

| Chit-CGO | | Chit-ECGO | |
|---|---|---|---|
| *Peak Position (eV)* | | *Peak Position (eV)* | |
| C1s | 285.0 | C1s | 285.0 |
| | 285.7 | | 286.7 |
| | 286.8 | | 288.4 |
| O1s | 532.3 | O1s | 532.4 |
| N1s | 398.8<br>400.3 | N1s | 397.7<br>399.7 |

The interplay between the defect structure of GO, polymer-induced supramolecular organization, and irradiation-driven transformations offers unique opportunities for controlling material morphology and physicochemical properties at micro- and nanoscales. SEM images of pristine Chit-functionalized CGO and ECGO films for 2, 5, 10, and 20 kV accelerating voltages are presented in **Figure S4**. As is obvious from **Figures S4(a-d)**, under the effect of the e-beam, structural variations are noted in Chit-CGO and depending on the accelerating voltage and exposure time we can form shell-like structures or microholes. However, as evident from **Figures S4(e–h)**, the irradiated regions in Chit-ECGO remain structurally intact regardless of exposure time, with no observable cracking or rupture of the formed holes. This behavior indicates that the dominant process is material densification rather than bubble-driven damage or ablation. **Figure S5** illustrates the response of Chitosan film with less than 10μm thickness under e-beam irradiation, demonstrating that the material reacts immediately even at an accelerating voltage of 5 kV. **Figure S6** shows e-beam-induced Chit-CGO, Chit-ECGO and Chitosan micropatterns created using different irradiation exposure times on the same film as well as on independent samples. The final morphology and properties of Chit-functionalized GO micropatterns heavily rely on the interplay between irradiation conditions and the chemical preparation of the precursor solution such as the pH, temperature and concentration of the composition. We have also performed AFM measurements of Chit-CGO and Chit-ECGO after electron-beam irradiation, which further confirm the formation of surface holes and protrusions (**Figure S7**).

A detailed liquid crystalline domain analysis was performed based on POM images for CGO, Chit-CGO, ECGO, and Chit-ECGO systems before and after e-beam irradiation indicating that, while the overall distribution across the cell is almost homogeneous, the domains remain partially localized (**Figure S8**).

### 3.4 Density Functional Theory (DFT) Analysis

To explore possible origins of the synthesis-dependent e-beam response, DFT calculations were performed on two simplified GO bilayer models: a highly oxidized configuration (CGO-like, $C_{64}H_7O_{14}$) and a moderately oxidized configuration (ECGO-like, $C_{64}H_5O_{10}$). These models isolate the effect of oxidation degree on structural stability and electronic distribution in an idealized setting, and they do not capture the full complexity of the experimental films.

#### 3.4.1 CGO-like system: Decarboxylation-driven hole formation

The optimized CGO model exhibits a high density of oxygen-containing functional groups, leading to significant local lattice distortion and non-uniform C-C bonding (**Figure S9(a)**). C–C bond length analysis reveals pronounced elongation in the vicinity of oxygen functionalities (1.504 Å) compared with less oxidized regions (1.420 Å), corresponding to ~5.9% bond weakening. This indicates that oxygen functional groups induce strong local strain which may facilitate bond rupture under irradiation [48]. To evaluate the thermodynamic feasibility of carbon loss, decarboxylation was modeled as GO → defective GO (hole) + $CO_2$. The calculated reaction energy is $E_{reaction} = -0.9992$ eV, see **Table S7**.

This negative value indicates that $CO_2$ release is thermodynamically allowed in the highly oxidized model. For comparison, the corresponding carbon-loss process in the ECGO-like model gives +0.62 eV, while oxygen-group removal from the CGO-like model without carbon loss gives +0.84 eV, indicating that carbon loss through decarboxylation is energetically more favorable in the CGO-like model.

However, a negative reaction energy alone does not establish decarboxylation as the dominant e-beam pathway. Beam damage is a non-equilibrium process involving electronic excitation and/or knock-on displacement at the applied e-beam energy, while reaction kinetics were not evaluated. We therefore regard decarboxylation as one plausible contributor to carbon loss and ablation in CGO, supported by the local C–C bond elongation, the vacancy-site reconstruction (**Figure S9(b))**, and the experimentally observed hole formation. Together, these results are consistent with the experimentally observed hole formation in CGO.

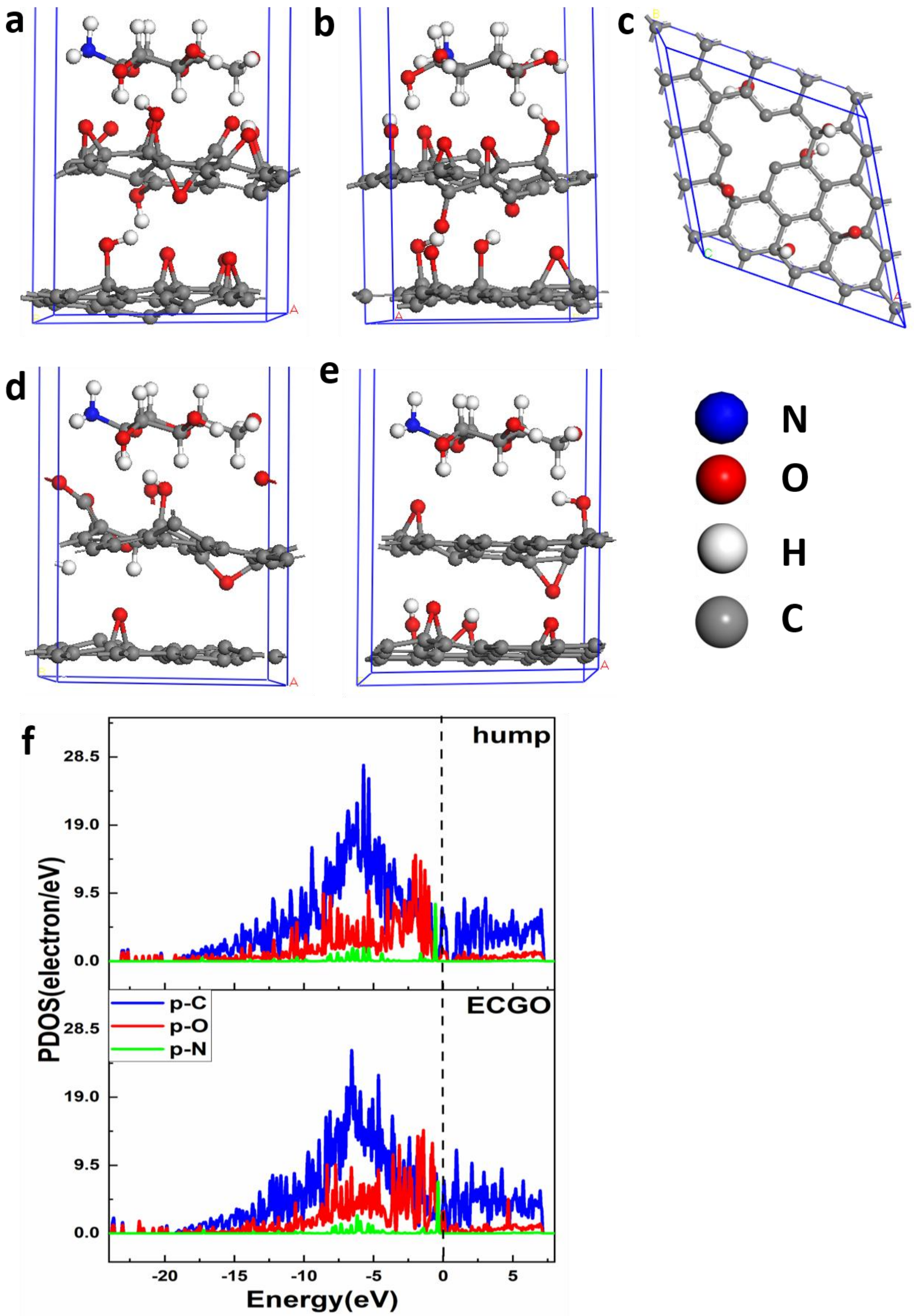

a
b
c
d
e
N
O
H
C
f
hump
ECGO
p-C
p-O
p-N
PDOS(electron/eV)
28.5
19.0
9.5
0.0
-20
-15
-10
-5
0
5
Energy(eV)

**Figure 6.** DFT analysis of chitosan-containing GO models. (a) Optimized structure of the chitosan-containing highly oxidized CGO-like model. (b) Optimized chitosan-containing CGO-like structure after carbon/oxygen removal, representing the proposed carbon-loss/ablation-related configuration. (c) Top view of the hole-defect structure highlighting the vacancy region and local lattice reconstruction. (d) Optimized structure of the chitosan-containing moderately oxidized ECGO-like model. (e) Optimized chitosan-containing ECGO-like structure after selective removal of oxygen-containing functional groups, showing the contracted/hump-like configuration. (f) Projected density of states (PDOS) of the chitosan-containing ECGO-like and reduced ECGO-like models. The vertical dashed line at 0 eV denotes the Fermi level.

### 3.4.2 ECGO-like system: Reduction-induced densification and hump formation

In contrast, the ECGO model ($C_{64}H_5O_{10}$) contains a lower concentration of oxygen functional groups and retains larger $sp^2$-conjugated domains (**Figure S9(c)**) [49]. Structural optimization shows that removal of oxygen functionalities leads to a significant contraction of the graphene oxide bilayer (**Figure S9(d))** [50].
The interlayer spacing decreases from:
$d=4.9\ Å \rightarrow 4.5\ Å$
corresponding to an ~8.2% reduction in layer separation, as shown in **Figure S10**. This contraction indicates enhanced interlayer π–π interactions and densification of the carbon framework upon reduction.

Charge redistribution analysis (Bader charge analysis) reveals that carbon atoms previously bonded to oxygen-containing groups exhibit increased electron density after oxygen removal. The most significant changes are observed for carbon atoms formerly bonded to hydroxyl groups ($\Delta q \approx +0.80$ to $+1.32$ e), indicating partial restoration of the conjugated $sp^2$ carbon network and electronic delocalization.

To further understand the electronic origin of this structural behavior, projected density of states (PDOS) analysis was performed **(Figure S9(e))**. The results show a significant suppression of oxygen-derived states and a corresponding enhancement of carbon 2p states near the Fermi level upon oxygen removal. This electronic reconstruction reflects partial recovery of the graphene-like π-network.

The combined structural and electronic results indicate that reduction does not lead to vacancy formation but instead promotes lattice stabilization, charge redistribution, and interlayer contraction. This densification process provides a direct atomistic explanation for the experimentally observed protrusion-like (hump) morphology in electrochemically exfoliated GO under e-beam irradiation.

## 3.4.3 Effect of a chitosan monomer on the GO models

Because the experimental materials are GO/chitosan composites rather than isolated GO sheets, the possible influence of chitosan was additionally examined by introducing a representative chitosan monomer onto the upper GO layer of both the CGO and ECGO models. The lowest-energy configurations obtained for the two systems are shown in Figure 6(a) and Figure 6(d), while the corresponding modified structures are presented in Figure 6(b) and Figure 6(e). The top view of the hole-defect structure, highlighting the vacancy region and the associated local lattice reconstruction, is presented in Figure 6(c).

The calculated interaction between chitosan and the GO surfaces is characterized by hydrogen-bond formation between the amino/hydroxyl groups of chitosan and oxygen-containing groups on GO with representative H-bond distances of approximately 1.75–1.82 Å. The calculated adsorption energies are approximately −0.82 eV for chitosan/CGO and −0.58 eV for chitosan/ECGO, respectively. These values indicate favorable local chitosan–GO interactions in both models. However, because a single monomer is used to represent chitosan and the present calculations do not explicitly describe the complete polymer network, these adsorption energies should be interpreted as qualitative indicators of local interfacial interaction rather than macroscopic composite binding energies.

For the chitosan-containing CGO model, the C–C bond near the oxygenated region changes from 1.504 to approximately 1.491 Å after chitosan adsorption. This slight contraction suggests that local chitosan–GO interactions can modify the electronic and structural environment around oxygenated carbon sites. The calculated energy for the corresponding carbon-loss/decarboxylation process in the presence of chitosan is provisionally estimated to be -0.42 eV, compared with −0.9992 eV for the isolated CGO model. This reduction in thermodynamic driving force suggests that interaction with chitosan may partially stabilize the oxygenated carbon environment and consequently reduce the energetic preference for the modeled carbon-loss process.

For the ECGO system, the oxygen-removal-associated structure retains the overall contraction observed in the GO-only model. The interlayer distance is approximately 4.55 Å in the chitosan-containing reduced ECGO structure, compared with approximately 4.50 Å for the corresponding GO-only model. Thus, the presence of the chitosan monomer slightly modifies, but does not eliminate, the contraction associated with the modeled oxygen-removal process.

The corresponding provisional reaction energy for oxygen removal from the chitosan-containing ECGO model is estimated as +0.30 eV, compared with the value obtained for the GO-only ECGO model. The qualitative comparison suggests that chitosan–GO interactions can modify the energetic cost of oxygen removal while preserving the overall structural tendency toward interlayer contraction.

The calculated charge-transfer values between chitosan and the GO surfaces are approximately 0.11–0.16 e, indicating modest interfacial electronic redistribution. The PDOS of the chitosan-containing ECGO and reduced ECGO structures (Figure 6(f)) shows contributions from N 2p and O 2p states mainly below the Fermi level, with limited hybridization close to the Fermi level. These results suggest that chitosan modifies the local electronic environment of GO without completely reconstructing the electronic states associated with the GO carbon framework.

Overall, the chitosan-containing calculations suggest that polymer–GO interactions can influence the local energetics of oxygenated groups and the interlayer geometry. In particular, the reduced driving force calculated for the modeled CGO carbon-loss process suggests a possible stabilizing effect of chitosan on oxygenated carbon sites. At the same time, the persistence of interlayer contraction in the ECGO model indicates that the structural response associated with oxygen removal is not eliminated by the presence of chitosan.

A single chitosan monomer cannot reproduce the full polymer chain, chain conformations, degree of polymerization, hydrogen-bond network, mechanical constraints, or experimentally used chitosan concentration. In addition, chitosan itself can undergo e-beam-induced chemical changes. Consequently, the present calculations provide local atomistic insight into chitosan–GO interactions, but they cannot quantitatively reproduce the complete radiation response of the composite.

### 3.4.4 Mechanistic interpretation: oxidation-dependent beam response

Altogether, the DFT results suggest an oxidation-dependent tendency in the e-beam response:

- In highly oxidized CGO, oxygen-induced bond elongation and thermodynamically allowed decarboxylation may promote carbon loss and vacancy growth, consistent with the ablation-driven hole formation observed experimentally.
- In moderately oxidized ECGO, oxygen removal stabilizes the carbon framework, enhances $sp^2$ restoration, and induces interlayer contraction, leading to densification and protrusion formation.

The chitosan-containing calculations further indicate that the polymer can modify the local energetic and electronic environment of both GO systems. In particular, the reduction in the calculated driving force for the CGO carbon-loss reaction in the presence of chitosan suggests that interfacial polymer interactions may partially stabilize oxygenated carbon sites. However, because the calculations employ simplified structural models and do not include explicit irradiation dynamics or transition-state barriers, they cannot uniquely determine the microscopic reaction sequence occurring under e-beam exposure.

Therefore, the DFT results are best interpreted as supporting evidence for oxidation-state-dependent degradation pathways: highly oxidized CGO is more compatible with carbon-loss/ablation-type evolution, whereas the less oxidized ECGO model is more compatible with oxygen-removal-associated structural contraction. The calculations complement, rather than replace, direct experimental characterization of elemental composition and chemical functionality after irradiation.

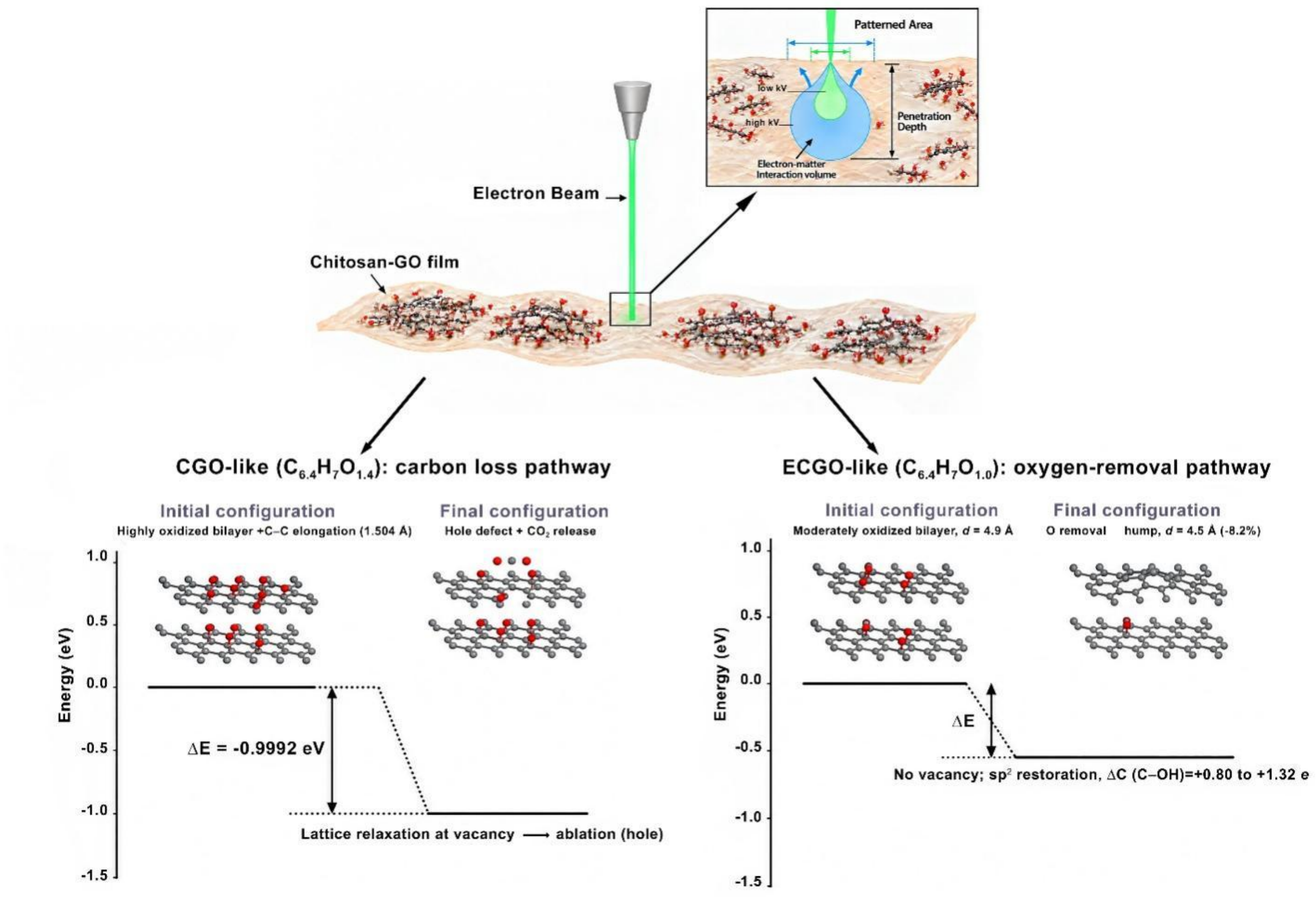

**Figure 7.** Figure 7. Electron-beam patterning of a chitosan–GO film via two pathways: (left) CGO-like composition undergoes carbon loss with $CO_2$ release and hole formation, while (right) ECGO-like composition undergoes oxygen removal, contracting the interlayer spacing from 4.9 to 4.5 Å.

## 4 Conclusion

In summary, GO was fabricated via both electrochemical and chemical exfoliation routes and afterward functionalized with Chit. The e-beam-induced morphological evolution of these two systems was systematically investigated, revealing distinct irradiation-driven dynamics depending on the GO synthesis pathway. XPS and Raman analyses provided comprehensive insights into the electronic and chemical states, and elemental composition of the constituent atoms, as well as the degree of $sp^2$ network disruption. SEM-based manipulation approach was demonstrated, enabling accurate regulation over µm-scale positioning of both Chit–CGO and Chit–ECGO systems. Furthermore, dose- and accelerating-voltage-dependent studies allowed the fabrication of Chit–GO micropatterns with precise geometry, feature size, inter-feature spacing, and spatial arrangement. The obtained findings establish a non-lithographic route for the direct patterning of Chit-GO and related hybrid structures. The DFT calculations provide atomistic insight into the different structural responses associated with the two GO oxidation states. For the highly oxidized CGO-like model, local C–C bond elongation and an exothermic decarboxylation endpoint support the plausibility of carbon-loss and vacancy formation. For the moderately oxidized ECGO-like model, oxygen removal is associated with charge redistribution, increased carbon-derived states near the Fermi level, and interlayer contraction. Additional chitosan-containing calculations indicate that Chitosan-GO interactions can modify, but do not necessarily eliminate, these local structural trends. Because the calculations represent selected model pathways and do not include irradiation kinetics or the complete polymer network, these results should be regarded as mechanistic support rather than definitive proof of a unique electron-beam pathway. Such controllable musk-free patterning enables their application in flexible electronics and biosensors, where the exfoliation method dictates the sensitivity and mechanical stability of the system.

# Supplementary Materials for

# Synthesis Method-dependent Resist-like Behavior of Liquid Crystalline Graphene Oxide Films Under Electron Beam Irradiation

A. Sahakyan[1], Ghada E. Khedr[2], A. A. Hovhannisyan[3], G. Gevorgyan[4], K. Margaryan[4], G. Shakhatuni[4], M. Aleksanyan[4], Mohamed Aly Saad Aly[5,6], N. Nwaji[7], H. Gharagulyan[1, 4*]

[1] *Liquid Crystalline Nanosystems Research Group, A.B. Nalbandyan Institute of Chemical Physics NAS RA, Yerevan 0014, Armenia*

[2] *Egyptian Petroleum Research Institute, Cairo 11727, Egypt*

[3] *The Scientific Technological Centre of Organic and Pharmaceutical Chemistry NAS RA, Molecule Structure Research Centre, 26 Azatutyan Av., Yerevan, 0014, Armenia*

[4] *Institute of Physics, Yerevan State University, Yerevan 0025, Armenia*

[5] *School of Electrical and Computer Engineering, Georgia Institute of Technology, Atlanta, Georgia 30332, United States*

[6] *Department of Electrical and Computer Engineering at Georgia Tech Shenzhen Institute (GTSI), Shenzhen, Guangdong 518055, China*

[7] *Institute of Fundamental Technological Research, Polish Academy of Sciences, Pawinskiego 5B Str. 02-106, Warsaw, Poland*

*Author to whom correspondence should be addressed: herminegharagulyan@ysu.am

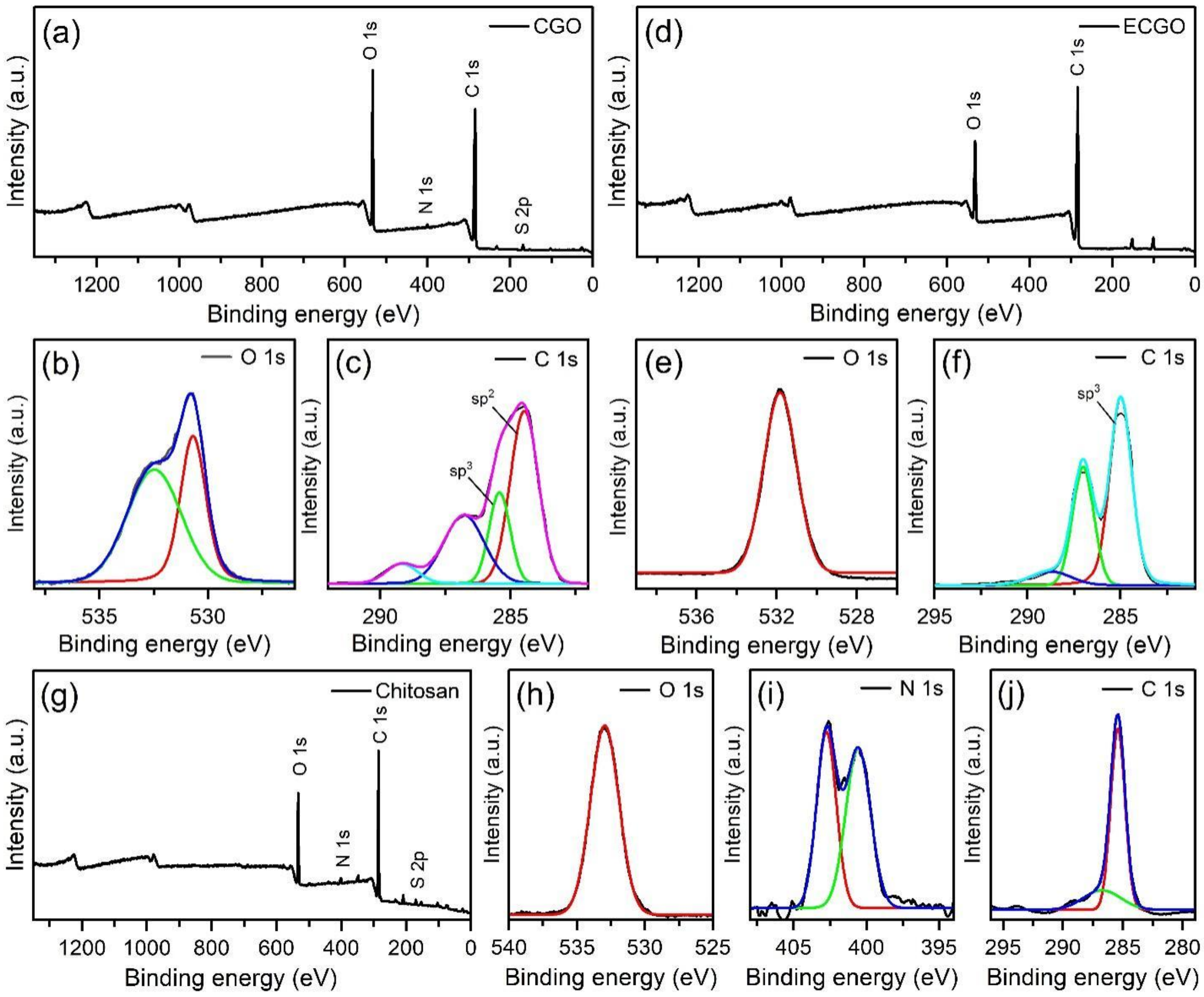

**Figure S1.** XPS spectra of CGO (a-c), ECGO (d-f) and pristine Chitosan (g-j).

**Table S1.** *R* values of pristine Chitosan, Chit-CGO, and Chit-ECGO composites.

| Samples | Chitosan | Chit-CGO | Chit-ECGO |
|---|---|---|---|
| *R* | 0.89 | 3.25 | 1.70 |

**Table S2.** A quantitative XPS comparison of the O/C ratio and deconvoluted functional groups for CGO, ECGO, Chit-CGO, and Chit-ECGO.

| Parameter | CGO | ECGO | Chit-CGO | Chit-ECGO |
|---|---|---|---|---|
| O/C | 0.60 | 0.32 | 0.55 | 0.43 |
| C–C/C=H (%) | 47.31 | 56.45 | 56.77 | 66.72 |
| C–OH, C–O–C (%) | 19.03 | - | 34.17 | - |
| –C=O (%) | 27.14 | 34.46 | - | 24.01 |
| O–C=O (%) | 6.52 | 9.09 | 9.06 | 9.27 |

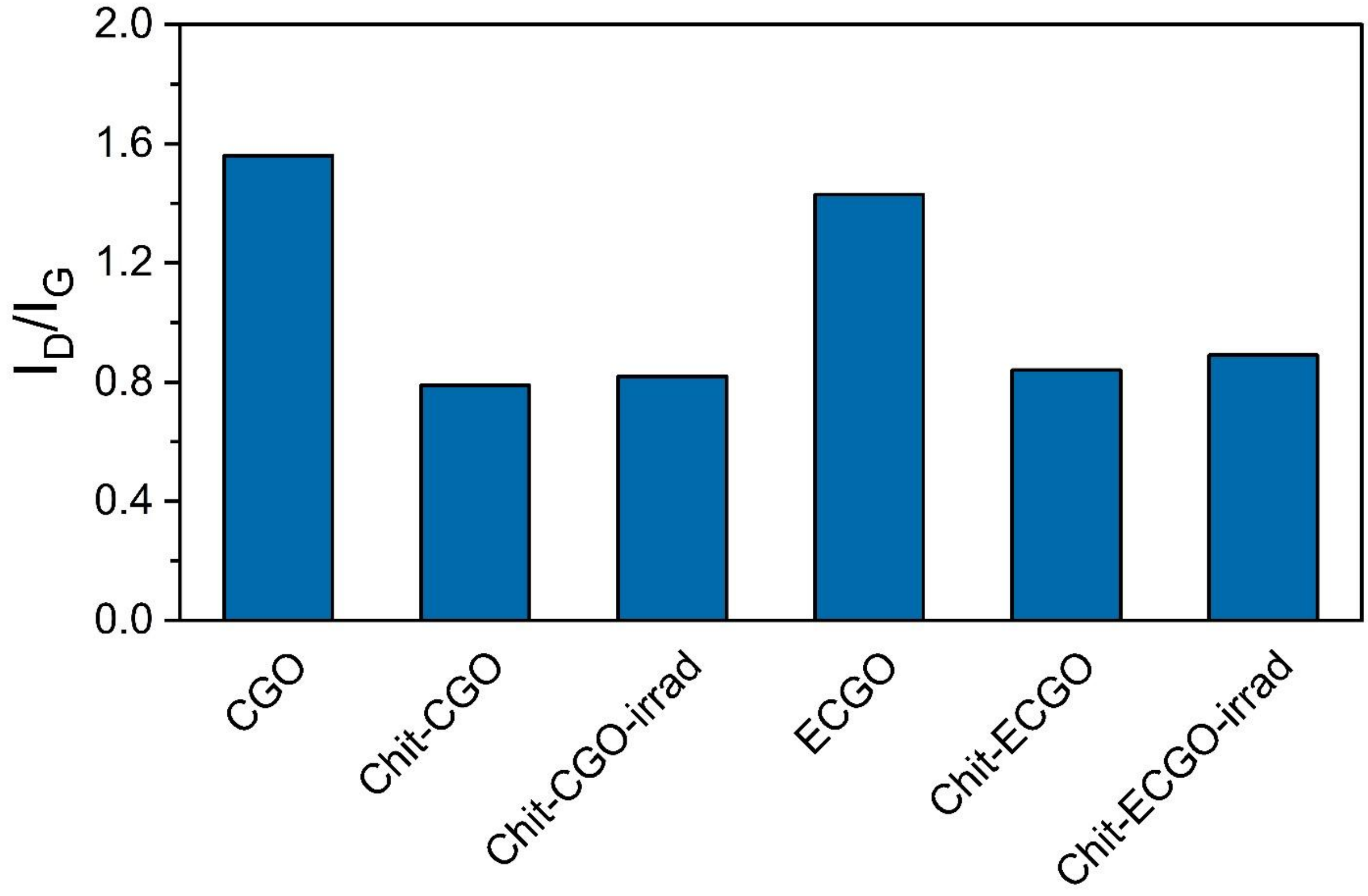

**Figure S2.** The $I_D/I_G$ intensity ratio of the bands of the tested materials.

**Table S3.** The values of *D* and *G* bands of Raman spectra for CGO, Chit-CGO, ECGO and Chit-ECGO systems.

| GO Sample | Curve | Peak Center (cm-1) | Peak area (arb. unit) | $I_D/I_G$ |
|---|---|---|---|---|
| CGO | D | 1343 | 654195 | 1.56 |
| | G | 1581 | 419356 | |
| Chit-CGO | D | 1340 | 220838 | 0.79 |
| | G | 1578 | 280724 | |
| Chit-CGO after irradiation | D | 1341 | 670957 | 0.82 |
| | G | 1580 | 816896 | |
| ECGO | D | 1350 | 745183 | 1.32 |
| | G | 1585 | 564532 | |
| Chit-ECGO | D | 1343 | 248074 | 0.84 |
| | G | 1578 | 294781 | |
| Chit-ECGO after irradiation | D<br>G | 1344<br>1564 | 643661<br>720313 | 0.89 |

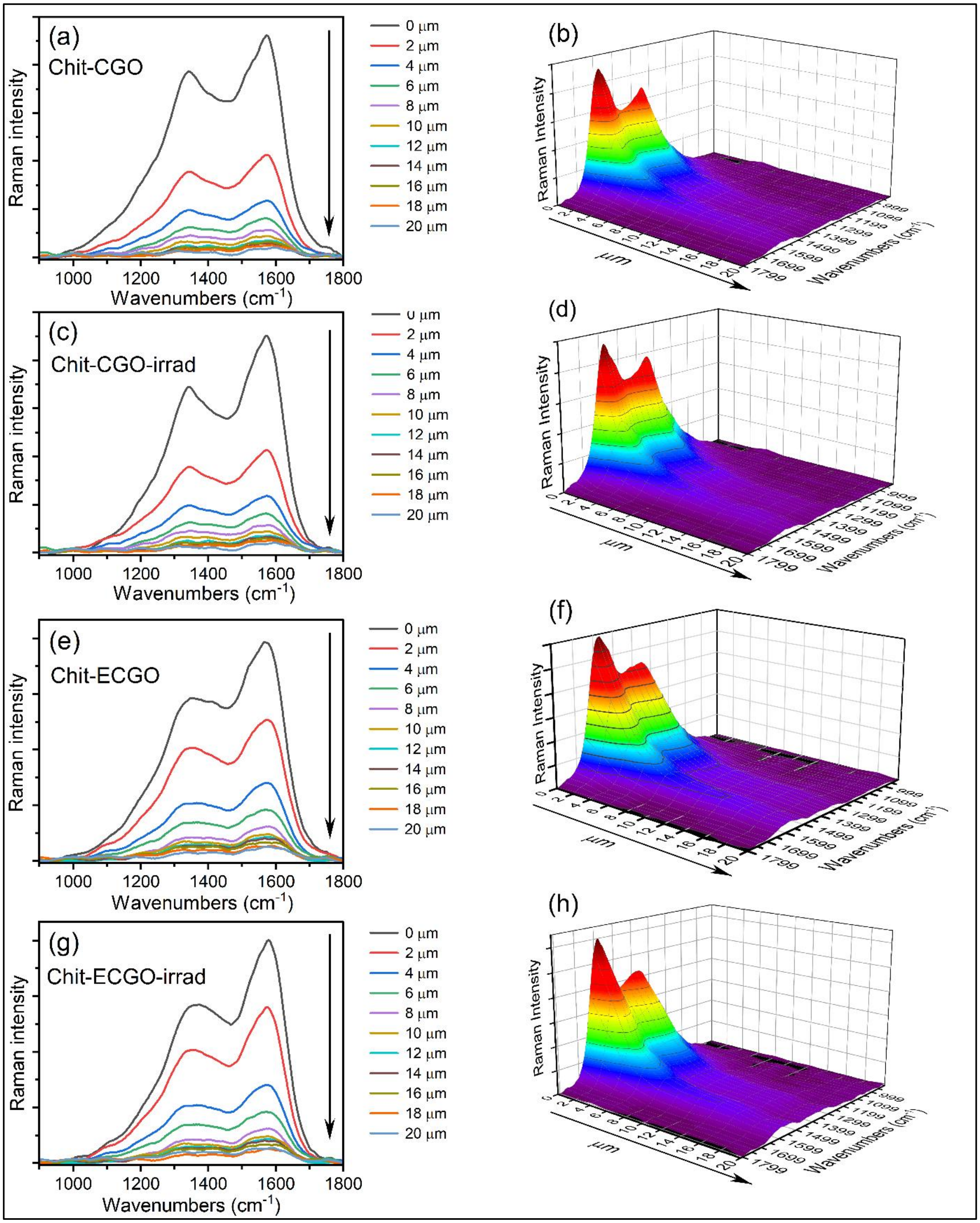

**Figure S3**. (a) Confocal Raman spectra of depth-profiling experiment, as a function of the probe *Z*-spatial position. The *Z*=0 position is 20 µm below the coverslip. The region of *D* and *G* bands are shown.

**Table S4.** Microhardness measurement data obtained at five selected points, together with the calculated average hardness values for Chitosan, CGO, ECGO, Chit-CGO, and Chit-ECGO films.

| Measurement Index | Result (HV 0.5/5) | | | | |
|---|---|---|---|---|---|
| | Chitosan | CGO | Chit-CGO | ECGO | Chit-ECGO |
| 1 | 12.54 | 31.04 | 7.43 | 133.37 | 11.84 |
| 2 | 5.01 | 48.01 | 11.51 | 110.80 | 9.52 |
| 3 | 6.84 | 27.29 | 12.13 | 104.46 | 8.62 |
| 4 | 10.02 | 51.99 | 10.45 | 95.34 | 10.74 |
| 5 | 9.62 | 32.61 | 8.31 | 95.10 | 9.69 |
| Mean | 8.80 | 38.18 | 9.96 | 107.81 | 10.08 |

**Table S5.** Absorbed irradiation doses and fluences by Chit-CGO, Chit-ECGO and Chitosan.

| **Irradiation** | **V (kV)** | **I (nA)** | **Area (cm$^2$)** | **Time (min)** | | | | |
|---|---|---|---|---|---|---|---|---|
| | | | | **1** | **2** | **3** | **5** | **10** |
| **Chit-CGO** | | | | | | | | |
| ***Dose (A·s/cm$^2$)*** | 10 | 0.31 | $1.74\times10^{-8}$ | 1.07 | 2.14 | 3.20 | 5.34 | 10.69 |
| ***Fluence (electron/cm$^2$)*** | | | | $6.68\times10^{18}$ | $1.34\times10^{19}$ | $2.00\times10^{19}$ | $3.34\times10^{19}$ | $6.68\times10^{19}$ |
| ***Dose*** | 5 | | $1.22\times10^{-8}$ | 1.52 | 3.05 | 4.57 | 7.62 | 15.25 |
| ***Fluence*** | | | | $9.53\times10^{18}$ | $1.91\times10^{19}$ | $2.86\times10^{19}$ | $4.76\times10^{19}$ | $9.53\times10^{19}$ |
| **Chit-ECGO** | | | | | | | | |
| ***Dose*** | 10 | 0.31 | $1.65\times10^{-8}$ | 1.13 | 2.25 | 3.38 | 5.63 | 11.27 |
| ***Fluence*** | | | | $7.05\times10^{18}$ | $1.41\times10^{19}$ | $2.11\times10^{19}$ | $3.52\times10^{19}$ | $7.05\times10^{19}$ |
| ***Dose*** | 5 | | $1.38\times10^{-8}$ | 1.35 | 2.70 | 4.04 | 6.74 | 13.48 |
| ***Fluence*** | | | | $8.42\times10^{18}$ | $1.68\times10^{19}$ | $2.53\times10^{19}$ | $4.21\times10^{19}$ | $8.42\times10^{19}$ |
| **Chitosan** | | | | | | | | |
| ***Dose*** | 10 | 0.31 | $2.42\times10^{-8}$ | 0.77 | 1.53 | 2.30 | 3.84 | 7.68 |
| ***Fluence*** | | | | $4.80\times10^{18}$ | $9.61\times10^{18}$ | $1.44\times10^{19}$ | $2.40\times10^{19}$ | $4.80\times10^{19}$ |
| ***Dose*** | 5 | | $0.52\times10^{-8}$ | 3.57 | 7.15 | 10.73 | 17.88 | 35.77 |
| ***Fluence*** | | | | $2.24\times10^{19}$ | $4.47\times10^{19}$ | $6.71\times10^{19}$ | $1.12\times10^{20}$ | $2.24\times10^{20}$ |

**Table S6.** Hole size, protrusion height, and feature dimensions as functions of electron-beam dose and film thickness.

| **Chit-CGO** | | | | |
|---|---|---|---|---|
| **Pattern features** | **Electron-beam dose (A·s/cm$^2$)** | | | |
| | 0.37 | | 1.52 | |
| ***Film thickness (µm)*** | 13.3 | 21.0 | 13.3 | 21.0 |
| ***Hole size (µm)*** | *d* = 4.7 | *d* = 4.1 | *d* = 5.4 | *d* = 4.9 |
| ***Hole depth (µm)*** | 0.30 | 0.17 | 0.31 | 0.26 |
| ***Inter-feature spacing (µm)*** | 10.12 | 11.34 | 7.70 | 9.23 |
| **Chit-ECGO** | | | | |
| **Pattern features** | **Electron-beam dose (A·s/cm$^2$)** | | | |
| | 0.31 | | 1.35 | |
| ***Film thickness (µm)*** | 3.8 | 4.3 | 3.8 | 4.3 |
| ***Protrusion size (µm)*** | 8.1×10.4 | 7.4×8.9 | 9.6×12.1 | 8.3×10.2 |
| ***Protrusion height (µm)*** | 0.70 | 0.62 | 0.82 | 0.68 |
| ***Inter-feature spacing (µm)*** | 4.2 | 5.3 | 3.7 | 4.8 |
| **Chitosan** | | | | |
| **Pattern features** | **Electron-beam dose (A·s/cm$^2$)** | | | |
| | 0.40 | | 3.57 | |
| ***Film thickness (μm)*** | 18.6 | 22.69 | 18.6 | 22.69 |
| ***Hole size (μm)*** | 8.1×5.8 | 8.4×6.3 | - | - |
| ***Hole depth (μm)*** | 0.18 | 0.20 | - | - |
| ***Inter-feature spacing (μm)*** | 1.57 | 1.34 | - | - |

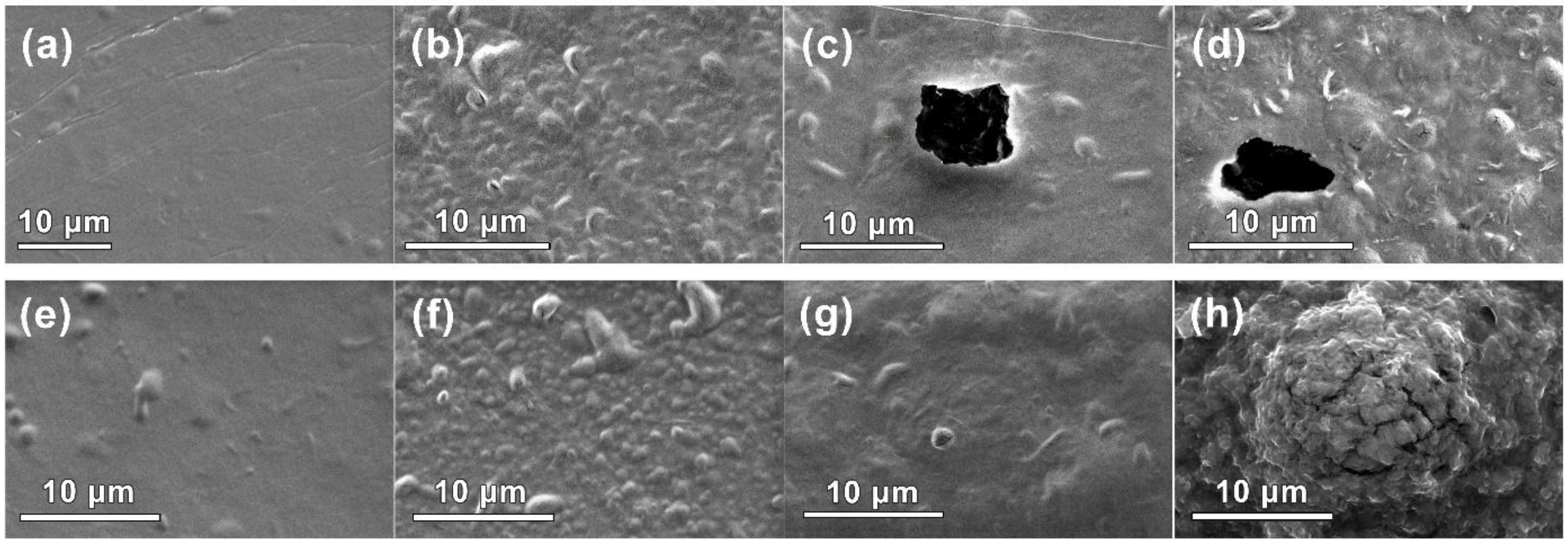

**Figure S4.** SEM images of pristine Chit-functionalized CGO (a, b, c, d) and ECGO (e, f, g, h) films for 2, 5, 10, and 20 kV accelerating voltages, respectively.

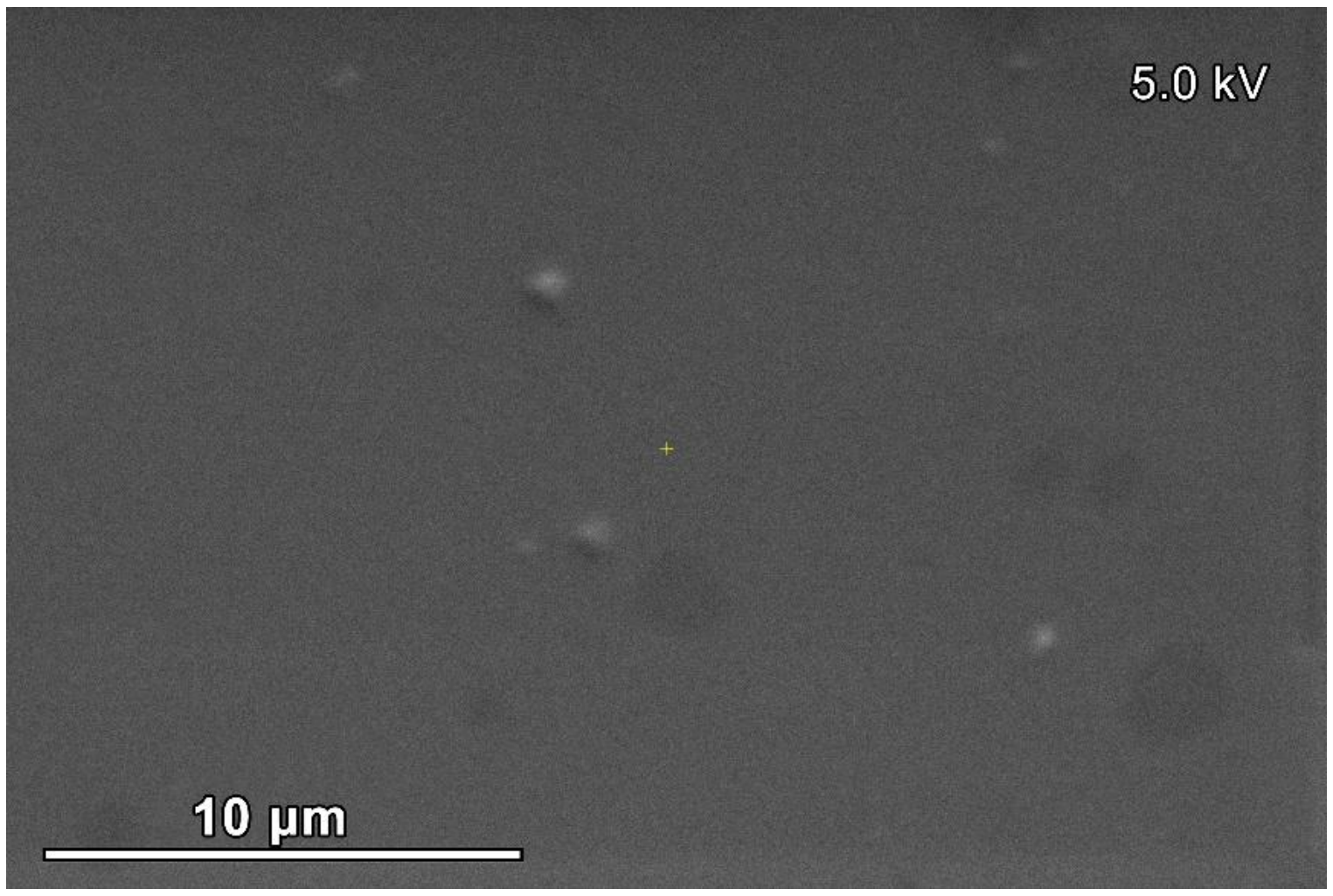

**Figure S5.** SEM image of Chitosan under 5 kV accelerating voltage.

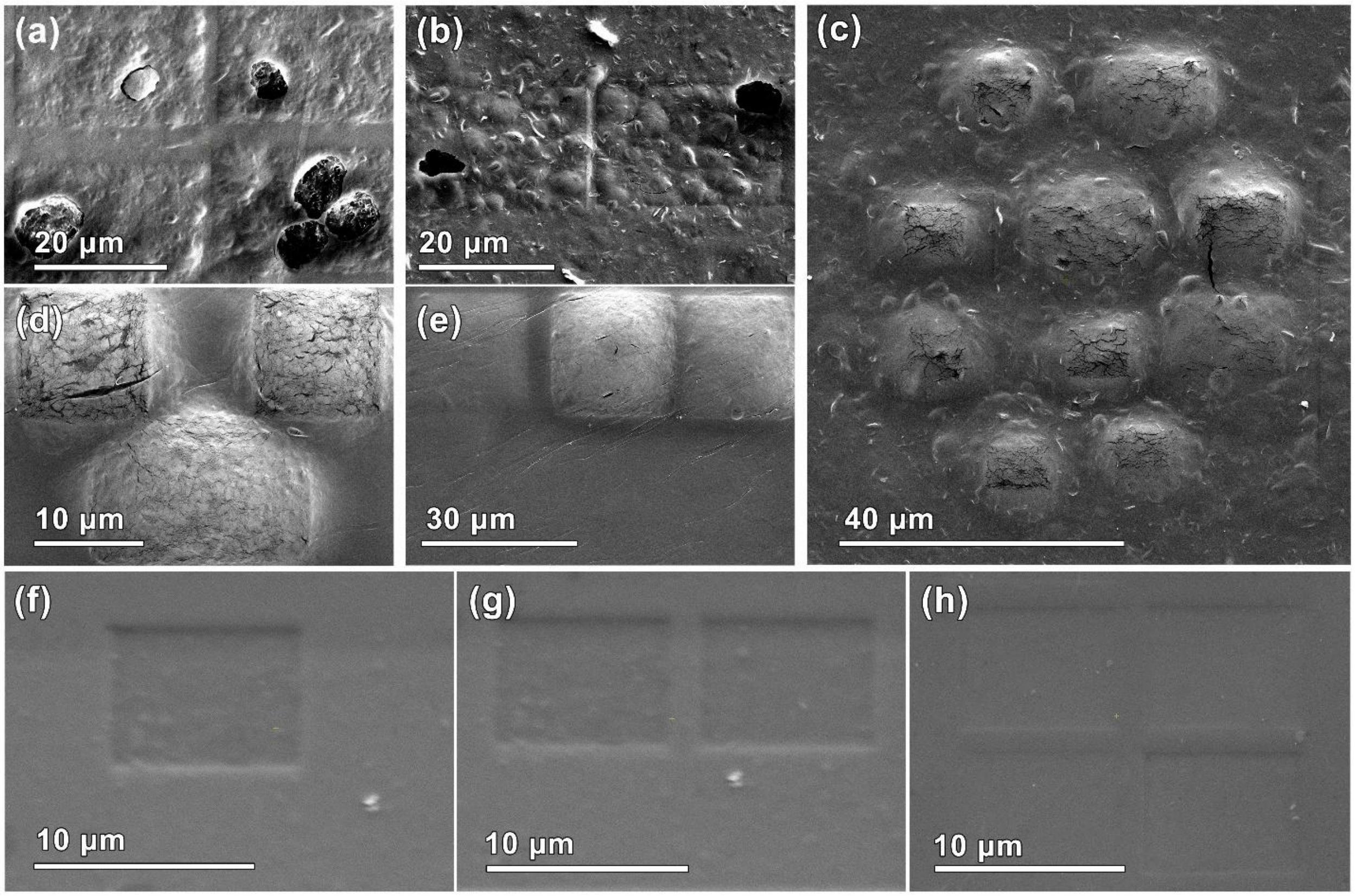

**Figure S6.** E-beam-induced micropatterns of Chit-CGO (a, b), Chit-ECGO (d, e), and Chitosan (f, g) created on independent samples and using different irradiation exposure times and doses. Panels (c) and (h) demonstrate e-beam-induced Chit-ECGO (c) and Chitosan (h) micropatterns obtained from multiple irradiated regions, demonstrating the reproducibility of the micropatterning process.

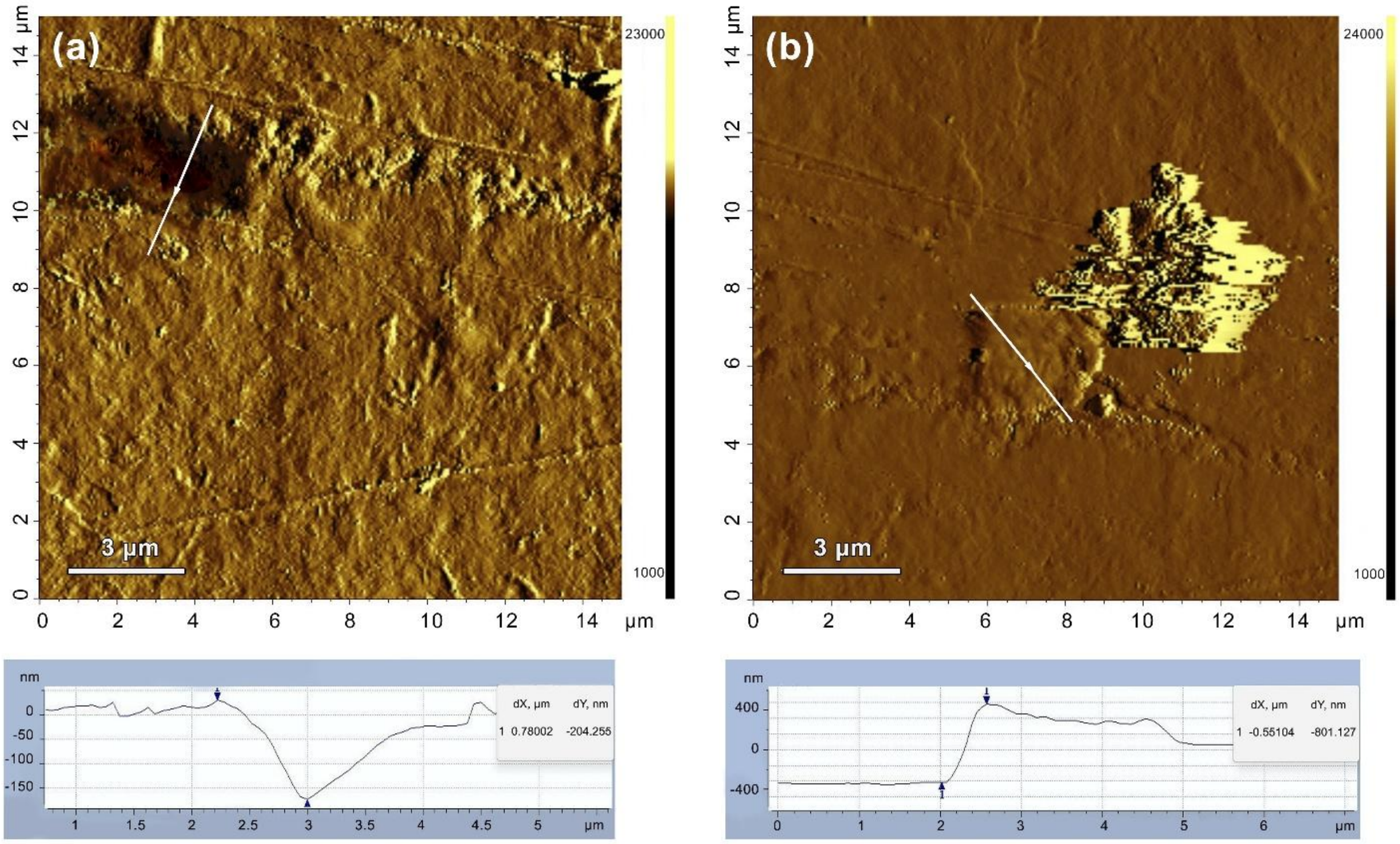

**Figure S7.** AFM images of (a) Chit-CGO and (b) Chit-ECGO after electron-beam irradiation.

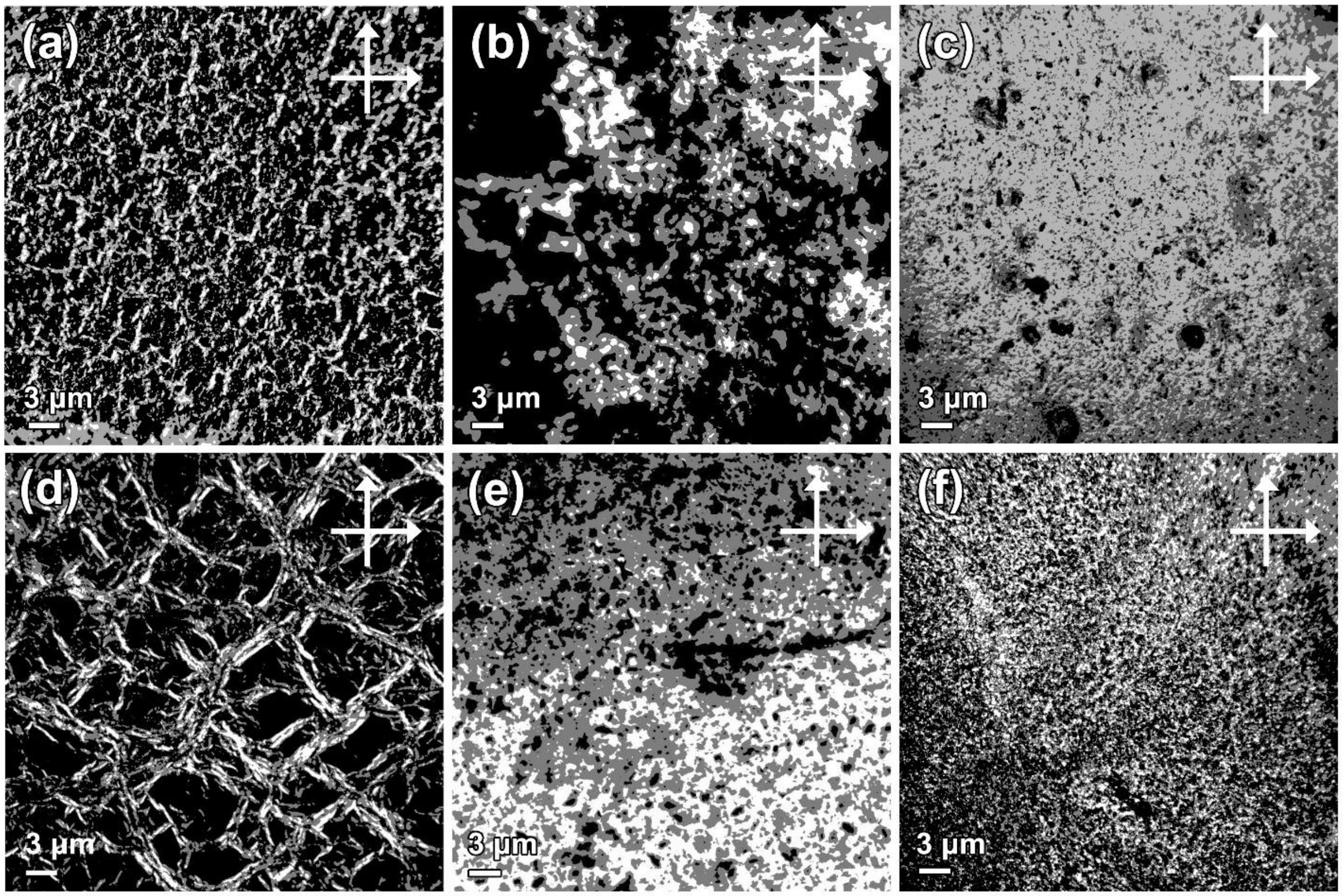

**Figure S8.** Liquid crystalline domain analysis of (a) CGO, (b) Chit-CGO, (d) ECGO and (e) Chit-ECGO POM images (b, e) before and (c, f) after e-beam irradiation: K-means clustering (k=3) segmented the image into dark, gray and bright regions. (a) Dark (64.25%), gray (27.32%), and bright regions (8.43%). (b) Dark (69.67%), gray (23.91%), and bright regions (6.42%). (c) Dark (40.86 %), gray (44.80 %), and bright regions (14.34 %). (d) Dark (63.49%), gray (30.06%), and bright regions (6.45%). (e) Dark (33.80%), gray (39.69%), and bright regions (26.51%). (f) Dark (57.10 %), gray (34.13 %), and bright regions (8.77 %). The higher percentage observed for the dark region is due to fewer clusters; for more precise estimation, a larger number of clusters should be considered. The samples examined after electron-beam irradiation had subsequently been coated with a thin Au layer for SEM characterization and it modifies the optical contrast of the POM images and consequently affects reliable grayscale-based segmentation and quantitative cluster analysis of the birefringent domains.

**Table S7.** Calculated total energies (in eV) for CGO, CGO-derived hole, ECGO, ECGO-derived hump, and $CO_2$ molecule.

| System | Energy (eV) |
|---|---|
| CGO ($C_{64}H_7O_{14}$) | -672.0855804 |
| CGO-derived hole | -650.105191 |
| ECGO ($C_{64}H_5O_{10}$) | -634.068684 |
| ECGO-derived hump | -627.7123197 |
| $CO_2$ | -22.97967954 |

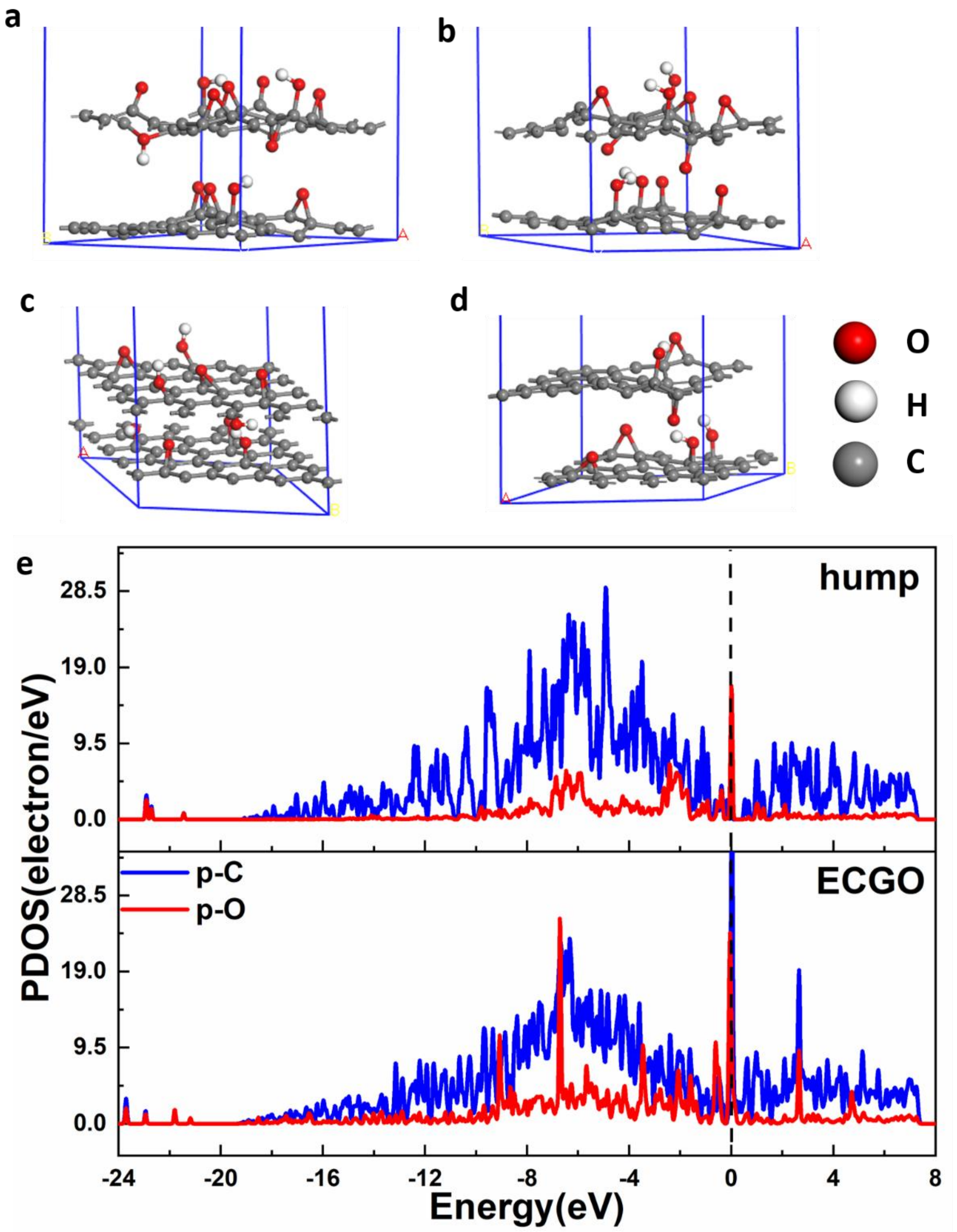

**Figure S9. (a)** Optimized structure of the highly oxidized chemically exfoliated graphene oxide (CGO) model. **(b)** Hole-defect structure obtained after carbon and oxygen removal from the CGO model, representing e-beam-induced ablation. **(c)** Optimized structure of the moderately oxidized electrochemically exfoliated graphene oxide (ECGO) model. **(d)** Reduced (hump) structure generated by selective removal of oxygen-containing functional groups from the ECGO model. **(e)** Projected density of states (PDOS) of the ECGO and hump structures. The vertical dashed line at 0 eV denotes the Fermi level.

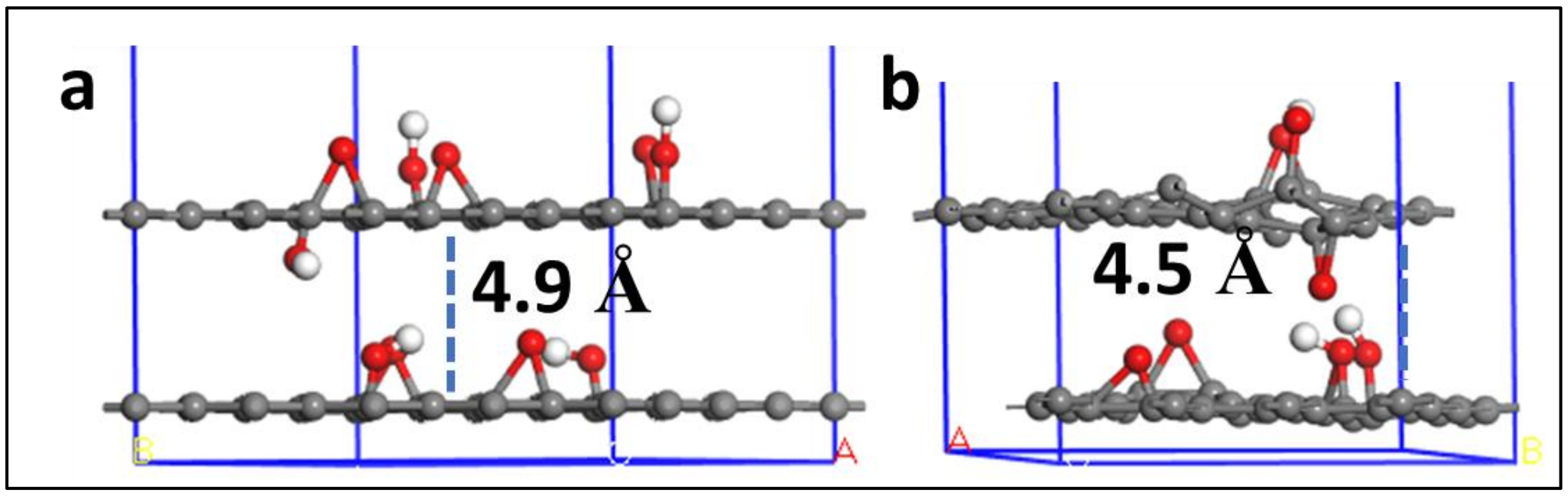

**Figure S10.** DFT-calculated reduction in interlayer spacing from 4.9 Å to 4.5 Å, corresponding to an approximately 8.2% decrease in layer separation.